\documentclass[5p,times,final,a4paper]{elsarticle}
\usepackage[utf8]{inputenc}
\usepackage{amssymb}
\usepackage{latexsym}
\usepackage{amsmath}
\usepackage{mathrsfs}

\usepackage[switch]{lineno}

\usepackage{siunitx}  
\usepackage{booktabs}
\usepackage{xcolor,soul,ulem}
\usepackage[hidelinks]{hyperref}
\hypersetup{
    colorlinks,
    linkcolor={red!50!black},
    citecolor={blue!50!black},
    urlcolor={black}
}

\usepackage{color}
\usepackage{enumitem}
\usepackage{ragged2e}

\journal{Computer Physics Communications}

\def\Section#1{Section~\ref{#1}}%
\def\Eq#1{Equation~\ref{#1}}%
\def\eq#1{Eq.~\ref{#1}}%
\def\Tab#1{Table~\ref{#1}}%
\def\tab#1{Tab.~\ref{#1}}%
\def\Fig#1{Figure~\ref{#1}}%
\def\fig#1{Fig.~\ref{#1}}%
\def\figs#1{Figs.~\ref{#1}}%
\def\Listing#1{Listing~\ref{#1}}%
\def\listing#1{Listing~\ref{#1}}%
\newcommand{\D}{\displaystyle}%
\newcommand{\thicktilde}[1]{\mathbf{\widetilde{\text{$#1$}}}}

\newcommand*{\Hbf}{\hat{\mathbf{H}}}
\newcommand*{\Vbf}{\hat{\mathbf{V}}}
\newcommand*{\Hbn}{\thicktilde{\mathbf{H}}}

\newcommand*{\sBN}{\vec{\thicktilde{\mathbf{s}}}}
\newcommand*{\wBN}{\vec{\thicktilde{\mathbf{w}}}}
\newcommand*{\vBN}{\vec{\thicktilde{\mathbf{v}}}}
\newcommand*{\upsiBN}{\vec{\thicktilde{\mathbf{\psi}}}}
\newcommand*{\HBN}{\thicktilde{\mathbf{H}}}
\newcommand*{\GBN}{\thicktilde{\mathbf{G}}}
\newcommand*{\GbarBN}{\bar{\thicktilde{\mathbf{G}}}}
\newcommand*{\UBN}{\thicktilde{\mathbf{U}}}
\newcommand\deff{\mathrel{\overset{\makebox[0pt]{\mbox{\normalfont\tiny\sffamily def}}}{=}}}
\newcommand\Iterate{\mathrel{\overset{\makebox[0pt]{\mbox{\normalfont\tiny\sffamily iterate}}}{~=~}}}

\usepackage{listingsutf8}
\usepackage{textgreek}

\definecolor{revAcolor}{HTML}{000000}
\newcommand{\revA}[1]{\textcolor{revAcolor}{#1}}

\definecolor{revBcolor}{HTML}{000000}
\newcommand{\revB}[1]{\textcolor{revBcolor}{#1}}

\definecolor{StyleColor}{HTML}{000000}
\newcommand{\revStyle}[1]{\textcolor{StyleColor}{#1}}

\begin{document}
\begin{frontmatter}
\title{%
%
%
%
%
%
%
Very accurate time propagation of coupled Schrödinger \revA{equations for} femto- and attosecond \revA{physics and} chemistry, with C++ source code
%
%
%
%
}
\author[gda1,gda2]{Janek Kozicki\corref{cor1}}
\address[gda1]{Faculty of Applied Physics and Mathematics, Gdańsk University of Technology, 80-233 Gdańsk, Poland}
\address[gda2]{Advanced Materials Center, Gdańsk University of Technology, 80-233 Gdańsk, Poland}
\cortext[cor1]{Corresponding author at Faculty of Applied Physics and Mathematics, Gdańsk University of Technology, 80-233 Gdańsk, Poland.}
\ead{jkozicki@pg.edu.pl}

\date{\today}

\begin{abstract}
\noindent
In this article, I present a very fast and high-precision
(up to 33 decimal places) C++ implementation of the
\textit{semi-global} time propagation algorithm for a system of coupled
Schrödinger equations with a time-dependent Hamiltonian.
It can be used to describe time-dependent processes
in molecular systems after excitation by femto- and attosecond laser pulses.
\revB{It also works with an arbitrary user supplied Hamiltonian}
\revA{and can be used for nonlinear problems}.
The \textit{semi-global} algorithm is briefly presented, the C++ implementation is
described and five sample simulations are shown. The accompanying C++ source
code package is included. The high precision benchmark (\texttt{long double}
and \texttt{float128}) shows the estimated calculation costs.
\revA{The presented method turns out to be faster and more accurate than the global Chebyshev propagator.}
\end{abstract}

\begin{keyword}
quantum dynamics \sep time dependent Hamiltonian \sep coupled Schrödinger equations \sep C++ \sep high precision \sep high accuracy
\noindent\rule{\textwidth}{0.4pt}
{\bf PROGRAM SUMMARY}~\\[-5mm]\onecolumn
\begin{small}
\begin{justify}
\begin{description}[leftmargin=!,labelwidth=47mm]
\item[Program Title:] SemiGlobalCpp
\item[CPC Library link to program files:] (to be added by Technical Editor)
\item[Developer's repository link:] \href{http://gitlab.com/cosurgi/SemiGlobalCpp}{\texttt{http://gitlab.com/cosurgi/SemiGlobalCpp}}
\item[Code Ocean capsule:] (to be added by Technical Editor)
\item[Licensing provisions:] GNU General Public License 2
\item[Programming language:] C++
\item[Operating system:] GNU/Linux
\item[Nature of problem:]
  The femto- and attosecond chemistry requires fast and high precision computation tools for quantum dynamics.
  Conventional software has problems with providing high precision calculation results (up to 33 significant digits), especially when the computation has to be as fast as possible.
\item[Solution method:]
  This software fills in the gap by providing the \textit{semi-global} \revA{algorithm}~\cite{Ndong2010,Schaefer2017,Schaefer2022} for arbitrary number of coupled electronic states for the time dependent Hamiltonian and nonlinear inhomogeneous source term. It is implemented in a way that allows computation with precision of 15, 18 or 33 significant digits, where the computation speed can be directly controlled by setting the required error tolerance.
  The \textit{semi-global} \revA{algorithm}~\cite{Ndong2010,Schaefer2017,Schaefer2022} is implemented in C++ providing a $10\times$ speed boost compared to original publication of \textit{semi-global} algorithm implemented in Matlab/\revA{Octave}~\cite{Ndong2010,Schaefer2017,Schaefer2022}.
\end{description}
\end{justify}
\end{small}~\\[-7mm]
\end{keyword}
\end{frontmatter}

\section{Introduction}
The time-dependent Schrödinger equation (TDSE):

\begin{equation}
\label{eqSchTDH}
i\hbar \frac{\partial\psi}{\partial t}=\Hbf(t)\psi,
\end{equation}

\noindent
is essential to quantum dynamics and especially to femto- and attosecond chemistry. The equation for a system of several coupled
time-dependent Schrödinger equations can be written down with explicitly shown
all $n$ coupled terms inside $\psi$ and $\Hbf(t)$:

\begin{equation}
\label{eqSchTDHCoupled}
i\hbar \frac{\partial}{\partial t}
\begin{pmatrix}
    {\psi}_1 \\[1mm]
    {\psi}_2 \\[1mm]
    \vdots   \\[1mm]
    {\psi}_n
\end{pmatrix}
=
\begin{bmatrix}
    \Hbf_1(t)     & \Vbf_{1,2}(t) & \dots  & \Vbf_{1,n}(t) \\[1mm]
    \Vbf_{2,1}(t) & \Hbf_2(t)     & \dots  & \Vbf_{2,n}(t) \\[1mm]
    \vdots        & \vdots        & \ddots & \vdots        \\[1mm]
    \Vbf_{n,1}(t) & \Vbf_{n,2}(t) & \dots  & \Hbf_{n}(t)
\end{bmatrix}
\begin{pmatrix}
    {\psi}_1 \\[1mm]
    {\psi}_2 \\[1mm]
    \vdots   \\[1mm]
    {\psi}_n
\end{pmatrix}
\end{equation}

\noindent
Each $\psi_n$ corresponds to a particular wavefunction at $n$-th electronic
potential, while $\Vbf_{j,k}$ correspond to the coupling elements between the
corresponding levels and $\Hbf_i(t)$ is the Hamiltonian at \revStyle{a} given level.
The solution to such system provides us with \revStyle{an} understanding of fundamental
quantum processes.  For almost all of these processes closed form solutions do
not exist.  Instead of these the numerical algorithms are being developed to
simulate the quantum processes from first principles.

This paper describes a general \textit{semi-global} algorithm for various classes of problems
including time-dependent Hamiltonian, nonlinear problems, non-hermitian
problems and problems with an inhomogeneous source \revA{term}~\cite{Ndong2010,Schaefer2017,Schaefer2022}.
%
%
The novelty is both the C++ implementation ($10\times$ faster than Octave) of this algorithm
as well as the addition of the ability to calculate multiple
coupled electronic levels in high precision (\texttt{long double} and
\texttt{float128} types having 18 and 33 decimal places respectively).

The main advantage of \revStyle{the presented algorithm} is the ability to significantly
reduce the numerical error in time propagation up to the level of the Unit in
the Last Place (ULP) error\textsuperscript{\ref{foot:ULP}} of the floating
point numerical representation (see Sections~\ref{sec:Calc:higher}
and~\ref{sec:High:Precision}, compare also with~\cite{Kahan2006}).  This is
made possible by
%
iteratively applying the Duhamel's principle until \revStyle{the} required convergence
criterion is met for $\Hbf(t)$. Additionally, the error tolerance can be
used to control the calculation speed.

This ability to produce results with errors in the range of the numerical ULP
error\footnote{\label{foot:ULP}ULP error for a given number $x$ is the smallest
distance $\epsilon$ towards the next, larger, number: $x+\epsilon$.%
} of used precision, together with high precision types \texttt{long double}
(18 decimal places) and \texttt{float128} (33 decimal places) and a fast C++
implementation means that the code presented in this work can be used to obtain
reference results in many difficult simulation cases.

This work is divided into \revStyle{the} following sections. In \revStyle{the} next section the theoretical
introduction to the \textit{semi-global} \revA{method}~\cite{Ndong2010,Schaefer2017,Schaefer2022} is presented.
In \Section{sec:Calc:higher} the high-precision calculations are briefly substantiated and described.
In \Section{chapter:Time:Dependent} the technical details of the C++ implementation are discussed.
In \Section{sec:validation} a validation of \revStyle{the} current implementation is performed.
In \Section{sec:High:Precision} the computation speed benchmarks are presented.
In \Section{sec:code:package} the accompanying C++ source code package is described and
finally, the conclusions are in \Section{sec:Conclusions}.

\section{The semi-global method}
\label{sec:SemiGlobalMethod}

The time-dependent Hamiltonian is useful for ultrafast spectroscopy, high
harmonic generation or coherent control problems.  In some cases the
Hamiltonian may become nonlinear depending explicitly on the state $\psi(t)$.
Such case occurs in mean field approximation, in the Gross-Pitaevskii
approximation~\cite{Bao2003}, time-dependent
Hartree~\cite{Balakrishnan1997,Beck2000,Kulander1987,Manthe2015} and
time-dependent DFT~\cite{Runge1984,Castro2006,Becke2014,Gross1990}. Another
complication may arise when adding a source term to the Schrödinger equation,
such as in scattering problems~\cite{Neuhauser1989}. All these cases will be
possible to calculate with the \textit{semi-global}~\cite{Schaefer2017,Schaefer2022,Tal2012}
method ported from Matlab/Octave~\cite{Schaefer2017,Schaefer2022} to C++ and described in this
paper.

The global scheme\footnote{,,global'' refers to the algorithm being independent
to the size of the timestep, contrary to typical Taylor based methods where the
power $n$ in $\Delta t^{n}$ refers to the order of the method. It treats
,,globally'' the entire time span of the calculation.} described in
\cite{Kosloff1984,Kosloff1997} assumes the knowledge of the eigenvalue range of the
Hamiltonian (i.e. the
$E_{min}$ and $E_{max}$). Usually, such knowledge is missing, especially for
time-dependent or non-Hermitian problems. To overcome this difficulty the
method below is implemented with the Arnoldi approach. The main advantage of
this approach is that the algorithm determines the energy range while
constructing the Krylov space.
A variation of \textit{semi-global} method also allows the Chebyshev
approach where the energy range is required~\cite{Schaefer2017}, this
approach is available in the attached C++ source code but is not discussed
with detail in this paper. For reference see~\cite{Schaefer2017,Schaefer2022}.

Several following sections summarize the detailed derivations presented
in~\cite{Schaefer2017} and thus are not a novelty in this work, however, they
are a crucial part to the final, novel,~\Section{sec:multiLevel} where the
method discussed here is extended to several coupled time-dependent Schrödinger
equations (such as \eq{eqSchTDHCoupled}). This final extension and its
novel high precision (up to 33 decimal digits) implementation in C++ allows to
simulate complex multi-level diatomic molecular systems. Another novelty is
that the presented C++ code is $10\times$ faster than the original Matlab/Octave
code \cite{Schaefer2017} (see \Section{sec:High:Precision}).


In other, less sophisticated, algorithms the method to overcome the Hamiltonian
time dependence is to use a small time\-step and assume constant Hamiltonian in
the single step. This becomes equivalent to \revStyle{the} method being first order in time
and there occurs a loss of precision which was gained by using a higher order
method. More sophisticated methods such as Magnus expansion~\cite{Magnus1954}
or high order splitting~\cite{Sun2012} do not assume stationary Hamiltonian,
but these methods are still local methods (they require relatively small
$\Delta t$) with a limited radius of convergence.

The method presented here is a \textit{semi-global} method, because it
combines the elements of local propagation and global propagation
methods~\cite{Tal2012}. A fully global time-dependent me\-thod was also
developed, but it turned out to be too expensive
computationally~\cite{Peskin1994}.  The \textit{semi-global} algorithm described here
is efficient with respect to accuracy compared to \revStyle{the} numerical effort.

\subsection{Establishing notation}

The time-dependent Schrödinger equation~\eq{eqSchTDH} is rewritten in a
discretized form:

\begin{equation}
\label{eqSchTDH1}
\frac{\partial\upsiBN(t)}{\partial t}=-\frac{i}{\hbar}\,\, \HBN(t)\upsiBN(t),
\end{equation}

\noindent
the time derivative of a discrete state vector $\upsiBN$ of finite
size\footnote{in the notation of $\upsiBN$ the tilde $\thicktilde\bullet$
indicates that it is discretized, while vector $\vec\bullet$ indicates that it
is a state vector.} is equal to a matrix (with time dependence) operating on
the same vector. Let us introduce a more general version of~\eq{eqSchTDH1} by
adding a source term $\sBN$ and by adding a dependence of $\HBN$ on
$\upsiBN(t)$:

\begin{equation}
\label{eqSchTDH2}
\frac{\partial\upsiBN(t)}{\partial t}=-\frac{i}{\hbar}\,\, \HBN(\upsiBN(t),t)\upsiBN(t) + \sBN(t).
\end{equation}

This is in fact a general set of ordinary differential equations (ODE). For
convenience let us define:\\[-6mm]

\begin{equation}
\label{eqSchTDH3}
\GBN(\upsiBN(t),t)\deff-\frac{i}{\hbar}\,\, \HBN(\upsiBN(t),t),
\end{equation}

\noindent
by incorporating the imaginary unit and Planck's constant into
$\GBN(\upsiBN(t),t)$. Thus we obtain the following set of ODEs to solve:

\begin{equation}
\label{eqSchTDH4}
\frac{\partial\upsiBN(t)}{\partial t}=\GBN(\upsiBN(t),t)\upsiBN(t) + \sBN(t),
\end{equation}
\noindent
and given the initial condition $\upsiBN_0\deff\upsiBN(t=0)$ the method
described in this section will allow propagation of the discrete state vector
to the next timestep $\upsiBN(t+\Delta t)$.

\subsection{Short summary for time-independent Hamiltonian}
\label{sec:TI:Ham}

%
%
\revA{The short summary in this subsection only serves the purpose of introducing
simpler versions of formulas which are
extended to time-dependent Hamiltonian in the following
subsections}\footnote{\revA{The content of this subsection is not implemented in the C++ code.}}.
\revA{The topics discussed in Sections~\ref{sec:TI:Ham}-\ref{sec:TDHam}
are about solving the same mathematical problem of approximating
${\vec{\mathbf{u}} = f(A)\vec{\mathbf{v}}}$. Both of them can be solved
by Polynomial expansion or Arnoldi,
where the Arnoldi algorithm is just a special case
of the polynomial approximation}.

We shall emphasize the main perk of all global time propagation methods: that
the evolution operator is expressed as a function of a matrix Hamiltonian
expanded over the energy spectrum using \revA{a chosen} polynomial basis, thus
allowing arbitrarily large timestep. Let us for a brief moment consider again a
simpler case without the time dependence and without the source term:\\[-6mm]

\begin{equation}
\label{eqSemi11}
\frac{\partial\upsiBN(t)}{\partial t}=\GBN_0\upsiBN(t),
\end{equation}

\noindent
where $\GBN_0\deff\GBN(t=0)$ loses time dependence. The evolution operator is
then expressed as an expansion in a truncated (to $K$ number of terms, see
\tab{tab:paramsKM} on page \pageref{tab:paramsKM} for a summary of
parameters in this method) polynomial series.
Hence the function $f(x)=e^{x \Delta t}$ ($\Delta t$ is a parameter) is
approximated as:\\[-6mm]

\begin{equation}
\label{eqSemi15}
f(x)\approx\sum_{n=0}^{K-1}a_n P_n(x),
\end{equation}
\noindent
where $P_n(x)$ is a polynomial of degree $n$ (e.g.~a Chebyshev polynomial, like
in~\cite{Kosloff1984}, but other polynomials can also be used here) and $a_n$ is
the expansion coefficient.  Then applying this evolution operator:
\begin{equation}
\label{eqSemi16}
\upsiBN(t+\Delta t)=e^{-\frac{i\,\,\HBN \Delta t}{\hbar}}\upsiBN\approx\sum_{n=0}^{K-1}a_n
P_n(\GBN_0)\upsiBN(t),
\end{equation}
\noindent
we obtain the solution\footnote{\label{foot:Scha:A}%
\revA{It should be clarified here, that the \eq{eqSemi16} (in which the operation of
the function of matrix on a vector uses a polynomial expansion approximation of}
\revA{the stationary evolution operator) and the first term in RHS of
Equations~\ref{eqSemi46} and \ref{eqSemi62} (in which the Arnoldi algorithm is
used to approximate a different function of a matrix which operates on a
vector) are actually both a special case of the solution of $\vec{\mathbf{u}} =
f(A)\vec{\mathbf{v}}$ which is a general problem in numerical analysis. The two
problems can be solved by the Arnoldi approach (see also \textit{Restarted
Arnoldi}~\cite{TalEzer2007}) or a polynomial expansion.  The Arnoldi algorithm
is a subtype of the polynomial expansion, because it is a polynomial
interpolation at estimated eigenvalues.
}}
at time $t+\Delta t$. The emphasis here lying on the
,,global'' property of the method: the error does not depend on the timestep
$\Delta t$. The solution is obtained directly at the final time $t+\Delta t$,
which can be arbitrarily large.  We shall note however that the
expansion~\eq{eqSemi16} has to be accurate in the eigenvalue domain of
$\GBN_0$.

We might consider the following polynomials $P_n$:

{\renewcommand{\labelenumi}{(\alph{enumi})}
\begin{enumerate}

\item Use the Taylor polynomials $P_n(x)=x^n$ and expand the evolution operator
in a Taylor series (a common approach in the local time integration methods),
but it is a poor choice: they are not orthogonal. To the contrary: as $n$
increases they are getting more and more parallel in the function space.

\item Use the Chebyshev polynomials $P_n(x)=T_n(x)$. The fact that they are
orthogonal to each other provides two useful properties: (1) the series
converges fast and (2) the expansion coefficients $a_n$ are given by a scalar
product of the $P_n(x)$ with $f(x)$. This approach is used
in~\cite{Kosloff1984}.

%

\item \revA{Use the Arnoldi approach which works well
when the spectral range of the Hamiltonian
is unknown. It uses the orthonormalized reduced Krylov basis representation.
This method will be summarized in \Section{sec:Arnoldi} and it is used in the
presented here \textit{semi-global} approach for time-dependent Hamiltonian}.

\end{enumerate}}

\noindent
We shall note that in~\eq{eqSemi16}, we obtain the solution only at the chosen
time $t+\Delta t$. It is desirable to follow the evolution of the physical
process at a smaller timestep, so that the time dependence of the Hamiltonian
\revB{(which is introduced in the following subsections)}
can be more accurately captured. It is possible to obtain these intermediate
time points $\Delta t_j\in[0,\Delta t)$ via negligible additional cost: using
the same matrix vector operations $P_n(\GBN_0)$ (Hamiltonian acting on the
wavefunction) but with different precomputed scalar coefficients $a_{n,j}$
(where $j$ corresponds to an intermediate time point in the evolution):\\[-8mm]

\begin{equation}
\label{eqSemi17}
\upsiBN(t+\Delta t_j)=e^{-\frac{i\,\,\HBN\,\,\Delta t_j}{\hbar}}\upsiBN_0\approx\sum_{n=0}^{K-1}a_{n,j} P_n(\GBN_0)\upsiBN_0\hspace{9mm}j=1, \dots , M, 
\end{equation}

\noindent
where $M$ is the number of additional intermediate time points in the solution.
It is possible, with low computation cost, because expansion of function in the
$P_n$ basis has different coefficients $a_{n,j}$ but has the same
$P_n(\GBN_0)\upsiBN_0$ evaluations.

The~\eq{eqSemi17} is in fact an expression for the evolution operator acting on
the wavefunction, which for the purpose of the next section we will denote as:

\begin{equation}
\label{eqSemi19}
\UBN_0(\Delta t_j)\deff e^{\GBN_0 \Delta t_j}.
\end{equation}

\noindent
Hence~\eq{eqSemi17} can be also written as:

\begin{equation}
\label{eqSemi17a}
\upsiBN(t + \Delta t_j)=\UBN_0(\Delta t_j)\upsiBN(t).
\end{equation}

\subsection{Source term with time dependence}

On our way towards the full time dependence in~\eq{eqSchTDH4}, we will now add
the source term with time dependence to~\eq{eqSemi11}:\\[-6mm]

\begin{equation}
\label{eqSemi18}
\frac{\partial\upsiBN(t)}{\partial t}=\GBN_0\upsiBN(t) + \sBN(t).
\end{equation}

We can integrate this equation using \revStyle{the} Duhamel principle, which provides a way to
go from the solution (\eq{eqSemi17}) of the homogeneous~\eq{eqSemi11} to the
solution of the inhomogeneous~\eq{eqSemi18} like this:

\begin{equation}
\label{eqSemi20}
\begin{split}
\upsiBN(t)&=\UBN_0(t)\upsiBN_0 + \int_0^t \UBN_0(t-\tau)\sBN(\tau)d\tau \\
&=e^{\GBN_0 t}\upsiBN_0 + \int_0^t e^{\GBN_0 t}e^{-\GBN_0 \tau}\,\,\sBN(\tau)d\tau\\
&=e^{\GBN_0 t}\upsiBN_0 + e^{\GBN_0 t}\int_0^t e^{-\GBN_0 \tau}\,\,\sBN(\tau)d\tau
\end{split}
\end{equation}


\noindent
Above we took the advantage of being able to extract from the Duhamel's
integral the $\UBN_0(t)$ part. So the only part which needs to be solved
explicitly is the remaining integral. To do this we will assume that
$\sBN(\tau)$ (after discretization: $\sBN(\Delta t_j)$) can be expressed as a
polynomial function of time. It is a bit of a simplification, but later on we
will be able to decide how many elements $M$ (see \tab{tab:paramsKM}) in
the series the algorithm will use, thus being able to directly control the
accuracy of the solution:\\[-6mm]

\begin{equation}
\label{eqSemi22}
\sBN(t)=\sum_{m=0}^{M-1}\frac{t^m}{m!}\sBN_m.
\end{equation}

\revA{It shall be noted here that a Chebyshev approximation of $\sBN(t)$ is used,
which is next converted to a Taylor form as in~\eq{eqSemi22}. By this way a much faster
convergence is achieved, which is an advantage of the \textit{local}
aspects of the \textit{semi-global} method}.

Together with the desired error tolerance
(to be introduced in~\Section{sec:TDHam}) and \revA{the parameter $K$
(to be introduced in~\Section{sec:Arnoldi})}
we have all the parameters which
govern the accuracy of the solution. The meaning of these parameters is
summarized in~\Section{sec:paramsKM}.

%
%

Let us now go back to calculating the integral present in \eq{eqSemi20}. Plugging~\eq{eqSemi22}
into~\eq{eqSemi20} yields:

\begin{equation}
\label{eqSemi34}
\upsiBN(t)=e^{\GBN_0 t}\upsiBN_0 + e^{\GBN_0 t}\sum_{m=0}^{M-1}\frac{1  }{m!}\int_0^t e^{-\GBN_0
\tau}\,\,\tau^m d\tau\,\, \sBN_m
\end{equation}

\noindent
Which with the following definitions of $f_m(\GBN_0,t)$, $\wBN_m$ and $\vBN_j$:

\begin{equation}
\label{eqSemi35}
f_m(z,t)\deff
\left\{
    \begin{array}{ll}
\D\frac{1}{z^m}\left(e^{z\,t}-\sum_{j=0}^{m-1}\frac{(z\,t)^j}{j!}\right)&\textrm{~~~~~for~~~~}z\neq0\\[2mm]
\D\frac{t^m}{m!}&\textrm{~~~~~for~~~~}z=0
    \end{array}
\right.
\end{equation}

\begin{equation}
\label{eqSemi37}
\wBN_m\deff
\left\{
    \begin{array}{ll}
\upsiBN_0&\textrm{~~~~~for~~~~}m=0\\
\sBN_{m-1}&\textrm{~~~~~for~~~~}0<m\leq M
    \end{array}
\right.
\end{equation}

\begin{equation}
\label{eqSemi47}
\vBN_j\deff\sum_{m=0}^{j}\GBN_0^{j-m}\wBN_m
\end{equation}

\noindent
following the derivation in~\cite{Schaefer2017} the solution can be written as:


\begin{equation}
\label{eqSemi46}
\upsiBN(t)=f_M(\GBN_0,t)\vBN_M+\sum_{j=0}^{M-1}\frac{t^j}{j!}\vBN_j,
\end{equation}

\noindent
where the $f_M(\GBN_0,t)$ is acting on $\vBN_M$ and the calculations are actually performed in the
discretized spectrum
$z\in\sigma(\GBN_0)$ (see \Section{sec:Arnoldi}). Now, since $\vBN_j$ satisfy the recurrence relation:


\begin{equation}
\label{eqSemi48}
\vBN_j=\GBN_0\vBN_{j-1}+\wBN_j,
\end{equation}

\noindent
the overall computational cost of~\eq{eqSemi46} is reduced to $M+K$ matrix
vector multiplications\footnote{\revA{For a direct polynomial approximation
it is $M+K-1$, however in the implemented C++ code the Arnoldi algorithm
is used (see~\Section{sec:Arnoldi}) for which an extra Hamiltonian
operation is necessary, hence it is $M+K$.}}.

\subsection{Introducing time-dependent Hamiltonian}
\label{sec:TDHam}

In the case of time-dependent Hamiltonian:\\[-5mm]

\begin{equation}
\label{eqSemi55}
\frac{\partial\upsiBN(t)}{\partial t}=-\frac{i}{\hbar}\,\, \HBN(t)\upsiBN(t) + \sBN(t)
\end{equation}

\noindent
or rather:\\[-5mm]

\begin{equation}
\label{eqSemi56}
\frac{\partial\upsiBN(t)}{\partial t}=\GBN(t)\upsiBN(t) + \sBN(t),
\end{equation}

\noindent
the Duhamel principle does not yield a closed form solution. Instead an
iterative procedure can be used to obtain better and better approximations of
the solution. First let us move the time dependence from $\GBN(t)$ to $\sBN$ by
defining an extended source term\footnote{%
\revA{%
It is called an ,,extended source term'' because it depends on the state
vector and is not a real source term in the strict sense.
}} $\sBN_e$:\\[-7mm]

\begin{equation}
\label{eqSemi59}
\sBN_e(\upsiBN(t),t)\deff\sBN(t)+\GbarBN(t)\upsiBN(t),
\end{equation}

\noindent
where $\GbarBN(t)\deff\GBN(t)-\GBN_{avg}$ and $\GBN_{avg}$ is average time-independent%
\footnote{\revA{%
The time averaging is done in a single timestep $\Delta t$, by performing
the evaluation of the Hamiltonian
in the middle time point of the timestep ($\lfloor M/2 \rfloor$
which corresponds to $\Delta t/2$), there is no costly averaging operation of any kind. See \Section{sec:NewtonInterpolation} for details
about how $M$ extra time points are added spanning the entire timestep $\Delta t$.
}}
component of
$\GBN$. The equation to be solved now has \revStyle{the} following form:\\[-7mm]

\begin{equation}
\label{eqSemi60}
\frac{\partial\upsiBN(t)}{\partial t}=\GBN_{avg}\upsiBN(t) + \sBN_e(\upsiBN(t),t).
\end{equation}

We can use the previous solution~\eq{eqSemi46} to approximate the time evolution of $\upsiBN(t)$:\\[-5mm]

\begin{equation}
\label{eqSemi62}
\upsiBN(t)\approx f_M(\GBN_{avg},t)\vBN_M+\sum_{j=0}^{M-1}\frac{t^j}{j!}\vBN_j,
\end{equation}

\noindent
where this time the $\vBN_j$ are computed from $\sBN_e$. It means that $\vBN_j$
depend on $\upsiBN(t)$ which is still unknown, however the solution can be
obtained via iterations. Upon first evaluation, we either extrapolate from
previous timestep $\Delta t$ (by putting $t+\Delta t$ into~\eq{eqSemi62}) or
when jump-starting the calculations we use $\upsiBN_0$. Next, in each
successive evaluation, we use the approximation from \revStyle{the} previous iteration within
the timestep $\Delta t$ (which spans $M$ \revA{time points}). We repeat the
iterative procedure until the convergence criterion at sub-step $M$ is met:\\[-5mm]

\begin{equation}
\label{eqSemiVII}
\frac{||\upsiBN_{new}-\upsiBN_{prev}||}{||\upsiBN_{prev}||} < \varepsilon.
\end{equation}

It means that this method has a radius of convergence \revStyle{that} directly depends on
the timestep $\Delta t$ covered in the single iteration, and \revStyle{contains} $M$
\revA{time points}. Too large $\Delta t$ will cause the successive iterations to
diverge, this is the reason why this method is not a fully global method but a
\textit{semi-global} method. The useful result of this situation is that one
can directly control the computation cost by setting an acceptable computation
error $\varepsilon$. For reference solutions it can be set to ULP numerical
precision, for faster calculations it can be a larger value. The novelty in
this work is that it works also for higher precision types such as \texttt{long
double} or \texttt{float128} with 33 decimal places
(\Tab{tbl:benchmarked:types}), thus enabling very accurate calculations.

Since we have put the dependence on $\upsiBN(t)$ into $\sBN_e(\upsiBN(t),t)$ it
is also computationally inexpensive to put this dependence into the Hamiltonian
hence the method described above works also for~\eq{eqSchTDH2}. So putting it
all together, the solution to~\eq{eqSchTDH2}:\\[-6mm]

\begin{equation*}
\label{eqSchTDH2again}
\frac{\partial\upsiBN(t)}{\partial t}=-\frac{i}{\hbar}\,\, \HBN(\upsiBN(t),t)\upsiBN(t) + \sBN(t).
\end{equation*}

\noindent
is following:\\[-9mm]

\begin{equation}
\label{eqSemi62a}
\upsiBN_{new}(t)\Iterate f_M(\GBN_{avg},t)\vBN_M(\upsiBN_{prev})+\sum_{j=0}^{M-1}\frac{t^j}{j!}\vBN_j(\upsiBN_{prev}),
\end{equation}

\noindent
where iterations are performed until the convergence condition~\eq{eqSemiVII} is
met. Thanks to being able to extrapolate $\upsiBN_{prev}$ into the next timestep
$\Delta t$ by putting $t+\Delta t$ into \revStyle{the} above equation and with a good choice of
$M$, $K$ parameters usually one iteration is enough to achieve desired
convergence.

\begin{table*}[t]
\caption{Summary of parameters of semi-global time propagation algorithm}%
\label{tab:paramsKM}
\begin{tabular*}{\textwidth}{l @{\extracolsep{\fill}} p{11cm} l}
\toprule
parameter                & meaning                                                                                                & equation          \\
\midrule
$K$                      & The number of expansion terms used for the computation of the function of matrix (evolution operator). &  $f_M(\GBN_{avg},t)\vBN_M$ in~\eq{eqSemi62a}$^\mathsection$    \\
$M$                      & The number of interior \revA{time points} in each timestep,~\Fig{fig:chebT}$^\dag$.                                  & \eq{eqSemi22}, \revB{\eq{eqSemi37}}     \\
$\varepsilon$            & Tolerance: the largest acceptable computation error.                                                   & \eq{eqSemiVII}    \\
$\Delta t$               & The length of the timestep interval (it contains $M$ \revA{time points spanning $\Delta t$}).                                   & \eq{eqSemi117mod} \\
\bottomrule
\end{tabular*}\\[-2mm]
\begin{flushleft}
$^\mathsection$~\footnotesize{\revA{See footnotes\textsuperscript{\ref{foot:Scha:A},\ref{foot:Scha:B}} for details. $K$ is defined as in~\Section{sec:TI:Ham}, but in the code the Arnoldi algorithm from~\Section{sec:Arnoldi} is used (with the same meaning of $K$).}}\\
$^\dag$~\footnotesize{\revB{$\sBN_e(\upsiBN(t),t)$ is expanded over $M$ time points hence it uses $M$ polynomial terms in the expansion to cover all time points inside the timestep.}}
\end{flushleft}
\end{table*}

\subsection{Chebyshev \revA{time points spanning $\Delta t$} and Newton interpolation}
\label{sec:NewtonInterpolation}



\begin{figure}[t]
\includegraphics[width=\columnwidth]{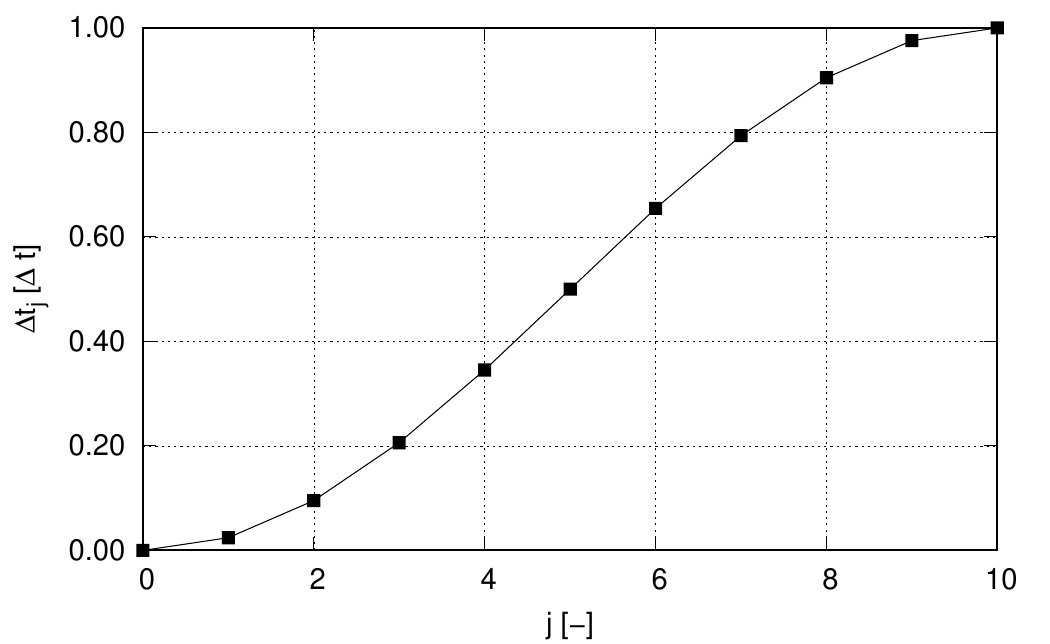}
\caption{Chebyshev time points \revA{spanning $\Delta t$} used for interpolation. They are equivalent to the
axis projection of points equally spaced on a unit semicircle.}
\label{fig:chebT}
\end{figure}

When interpolating a function \revA{(in this case it is a \textit{vector}
approximation of $\sBN_e$ at the time points)}, an equidistant set of
points is not a good choice: the closer to the boundary of the interpolation
domain the less accurate is the interpolation. This effect is known as the
Runge phenomenon. A much better set of points is with points becoming denser
closer to the edges of the domain.
\revA{It is called the Chebyshev sampling and it is required to perform the Chebyshev
approximation of $\sBN_e$ in time}.
Such a set of points \revA{is} chosen with \revStyle{the} following
formula:\\[-6mm]

\begin{equation}
\label{eqSemi117mod}
\revA{\Delta t_j\deff \frac{1}{2}\left(1-\cos\left(\frac{j\,\pi}{M-1}\right)\right)\Delta t,}
\end{equation}

\noindent
see example for M=11~in \fig{fig:chebT}.
These points are used as the
\revA{time points $\Delta t_j$, to interpolate $\sBN_e$, spanning the whole single} $\Delta t$ discussed in preceding sections.
%
\revA{The Chebyshev approximation mentioned here is different from the one
discussed in \Section{sec:TI:Ham} which approximates the function of a matrix
that operates on a vector.}

A Newton interpolation of function $f(t)$ at points $\Delta t_j$ is defined as:\\[-7mm]

\begin{equation}
\label{eqSemi107}
f(t)\approx \sum_{n=0}^N a_n R_n(t)
\end{equation}

\noindent
where $a_n$ are coefficients of the expansion and $R_n$ are the Newton basis
polynomials defined as: $R_0(t)=1$ and $R_n(t)=\prod_{j=0}^{n-1}(t-\Delta
t_j)$. \revB{During the process of calculating the Newton interpolation of
$f(t)$ the $a_n$ coefficients of the expansion are calculated as the divided
differences of the interpolated function $f(t)$ (see~\cite{Abramowitz2013}
sections 25.1.4 and 25.2.26 on pages 877 and 880; also see~\cite{Schaefer2017}
appendix A.1)}. It is used in~\eq{eqSemi15}.

\subsection{Arnoldi approach}
\label{sec:Arnoldi}

Calculation of the evolution operator in~\cite{Kosloff1984} method requires the
knowledge of \revStyle{a} spectral range of this operator. That method cannot be used when
it is impossible to estimate the eigenvalue domain. The difficulty with such
estimation grows when the eigenvalues are distributed on the complex plane,
which is the case with absorbing boundary conditions
(see~\Section{sec:dampingComplex}). And almost all interesting use cases
of time propagation (e.g. a multi-level diatomic molecular system evolving
under a laser impulse) require absorbing boundary conditions.  In such, quite
common, situation the Arnoldi approach comes to the rescue, because it works
without required prior knowledge of the eigenvalue domain.

The Arnoldi approach calculates the $f_M(\GBN_{avg},t)\vBN_M$ in~\eq{eqSemi62a}
in \revStyle{the} following manner:

{\renewcommand{\labelenumi}{(\alph{enumi})}
\begin{enumerate}

\item First construct an orthonormalized (via Gram-Schmidt process) reduced Krylov subspace $\vBN$,
$A\vBN$, $A^2\,\vBN$, $\dots$ , \revA{$A^{K-1}\,\vBN$} (the $K$ parameter controls the accuracy,
see~\Section{sec:paramsKM} and $A=\GBN_{avg}$)\footnote{\label{foot:Scha:B}%
\revA{The Arnoldi algorithm is a subtype of a polynomial expansion
with the number of terms being the size of the Krylov space, hence the parameter
$K$ of the Arnoldi method has the same meaning as in~\eq{eqSemi15}}.}.

\item Construct the transformation matrix $\Upsilon$ from the reduced Krylov basis representation to
the position representation of $\vBN_j$ vectors.

\item \revA{Rescale the eigenvalue domain during} the process using method
\cite{TalEzer1989} by dividing by the \textit{capacity of the domain} to reduce numerical errors.

\item Perform the calculation in the reduced Krylov basis representation then transform the result
back to original positional representation of $\vBN_j$ using the transformation matrix
$\Upsilon$.

\end{enumerate}}

\subsection{Extension to coupled time-dependent Schrödinger equations}
\label{sec:multiLevel}


The critical observation to extend this \textit{semi-global} algorithm to an arbitrary
number of coupled Schrödinger equations (see \eq{eqSchTDHCoupled}) is that this
method is independent of the Hamiltonian used. There is no requirement that
this Hamiltonian is a single-level system or several coupled levels.
The method uses the iterative procedure
to obtain the solution. The iterations are performed via invoking the $\Hbf(t)$
acting on $\psi$ until a convergence criterion is met: the difference between
the wavefunction from previous iteration and current iteration is smaller than
predetermined tolerance $\varepsilon$. This iterative procedure is executed inside
a single timestep $\Delta t$ over the $M$ \revA{time points}.

During this calculation all the expansion coefficients, both with respect to
$K$ (\eq{eqSemi15}) and with respect to $M$ (\eq{eqSemi22}) are stored inside a
matrix of size $N\times K$ and $N\times M$ respectively, where $N$ is the
number of grid points in the discretized wavefunction. The usual matrix algebra
is used in the C++ implementation, hence preserving the original code of the
implementation when moving to several coupled Schrödinger equations is
desirable. And it is possible because the algorithm itself is agnostic to the
Hamiltonian: it only invokes the $\Hbn(t)\psi$ in the computation. The trick
lies in the memory layout of the storage: the coupled wavefunctions are stored
one after another inside a single-column vector. A system of $k$ coupled
wavefunction uses $k\,N$ grid points. The system simply becomes larger and the
\textit{semi-global} algorithm uses matrices of sizes $(k\,N)\times K$ and $(k\,N)\times
M$ respectively without being informed how the information (stored in the
column vector of size $k\,N$) is used by the Hamiltonian.  The technical
implementation details are discussed in
\Section{sec:Coupled:Implementation}.  Additionally, the same approach
can be used when adapting \textit{semi-global} algorithm to solving higher dimensional
systems with more spatial directions, such as a system of three atoms using
Jacobi coordinates and with coupled Schrödinger equations therein.


\subsection{Summary of parameters used by the semi-global method}
\label{sec:paramsKM}

The \textit{semi-global} time integration method uses following parameters: $K$
introduced in~\eq{eqSemi16} to compute the evolution operator, $M$ introduced
in~\eq{eqSemi22} to help with the convergence of the iterative process, $\varepsilon$ -- the
error tolerance (\eq{eqSemiVII}) and the ,,global'' timestep $\Delta
t$ (\eq{eqSemi117mod}) over which the converging sub-iterations are being
computed. The short summary of these parameters is listed
in~\Tab{tab:paramsKM}.

\subsection{Absorbing boundary conditions with a complex potential}
\label{sec:dampingComplex}

An optimized complex absorbing potential is
used~\cite{Schaefer2017,Muga2004,Palao1998}. In the damping band\footnote{\revB{The
damping band is a region of space near the boundaries of the simulated
grid in which damping occurs. Usually, its width is several atomic units of length,
though it always has to be chosen carefully to fit the current simulation, because
too narrow band will result in \revStyle{the} reflection of the wavefunction and too wide band
(requiring larger grid) will be a waste of computer resources.}}
a sequence of
square complex barriers \revStyle{are} placed, each of them having their own reflection and
transmission amplitudes for a plane wave~\cite{Palao1998}. The parameters of
each barrier are optimized with respect to the cumulated plane wave survival
probability of all barriers. The typical characteristic of such \revStyle{a} barrier is that
it damps momentum within a certain momentum range and this range pertains to
the current problem being calculated. After the optimization procedure is
complete the complex potential is added to the potential used in the time
propagation. The Arnoldi approach works correctly with complex potentials (see
~\Section{sec:Arnoldi}) and this is why it is used by the \textit{semi-global} time
propagation method.


\section{Calculations in higher numerical precision}
\label{sec:Calc:higher}


%


There is a need for higher precision computations in quantum dynamics,
especially for attosecond laser impulses where the simulated time span is very
short and the timestep is very small~\cite{Palacios2019}. A very high temporal
resolution is necessary to describe such \revStyle{a} system. It stems from the simple fact
that with small timesteps (which are necessary to describe a quickly changing
electromagnetic field) there are a lot of small contributions from each
timestep. Upon adding these contributions many times, the errors will
accumulate\textsuperscript{\ref{precision_benefits}}. And only higher precision
calculations can make the errors significantly smaller.

High-precision
computation is an attractive solution in such situations, because even if a
numerically better algorithm with smaller error or faster convergence is known
for a given problem (e.g.~Kahan summation~\cite{Goldberg1991} for avoiding
accumulating errors\footnote{\label{precision_benefits}for $n$ summands
and $\varepsilon$ Unit in Last Place (ULP) error, the error in regular
summation is $n\varepsilon$, error in Kahan summation~\cite{Goldberg1991} is
$2\varepsilon$, while error with regular summation in twice higher
precision is $n\varepsilon^2$. See proof of Theorem
8~in~\cite{Goldberg1991}.}), it is often easier and more efficient to
increase the precision of an existing algorithm rather than deriving
and implementing a new one~\cite{Bailey2012,Kahan2004}.
%
However, switching to
high-precision generally means longer run
times~\cite{Isupov2020,Fousse2007} as shown in the benchmarks in
\Section{sec:High:Precision}.

Nowadays, high-precision computing is used in various fie\-lds, such as quantum
dynamics for physical and chemical purposes~\cite{Warren1993,Sakmann2011,Pachucki2005,Silkowski2020},
%
long-term stability analysis of the solar
system~\cite{Laskar2009,Sussman1992,Bailey2012}, supernova
simulations~\cite{Hauschildt1999}, climate modeling~\cite{He2001}, Coulomb
n-body atomic simulations~\cite{Bailey2002,Frolov2004}, studies of the fine
structure constant~\cite{Yan2003,Zhang1996}, identification of constants in
quantum field theory~\cite{Broadhurst1999,Bailey2005a}, numerical integration
in experimental mathematics~\cite{Bailey2005b,Lu2010}, three-dimensional
incompressible Euler flows~\cite{Caflisch1993}, fluid undergoing vortex sheet
roll-up~\cite{Bailey2005a}, integer relation detection~\cite{Bailey2000},
finding sinks in the Henon Map~\cite{Joldes2014} and iterating the Lorenz
attractor~\cite{Abad2011}. There are many more yet unsolved high-precision
problems in electrodynamics~\cite{Stefanski2013}.
In quantum mechanics the extended Hylleraas three electron integrals are
calculated with 32 digits of precision in~\cite{Pachucki2005}.
The long range asymptotics of exchange energy in a hydrogen molecule is
calculated with 230 digits of precision in~\cite{Silkowski2020}.
In quantum
field theory calculations of massive 3-loop Feynman diagrams are done with 10000
decimal digits of precision in~\cite{Broadhurst1999}.
Moreover the experiments in CERN are being performed in higher and higher precision
as discussed in the ``Welcome to the precision era'' article~\cite{CERNprecision}.
It brings focus to precision calculations and measurements which are performed
to test the Standard Model as thoroughly as possible, since any kind
of deviation will indicate a sign of new physics.



\revStyle{Consequently,} I believe that in the future, high-precision calculations will also
become more popular and necessary for the development of femto- and
attosecond chemistry. \revA{Following two use cases come to mind:}
{\renewcommand{\labelenumi}{(\alph{enumi})}
\begin{enumerate}
\item \revA{In order to properly describe a rapidly changing electric field of
a laser impulse in a typical femtosecond and attosecond chemistry problem a
small timestep is necessary~\cite{Tannor2007}. In such situation the
accumulation of machine errors can sometimes be a problem which can be readily
solved by high precision.}
\item \revA{High precision offers great resolution in the spectrum obtained from the
autocorrelation function which is commonly used to numerically identify the
oscillation energy levels~\cite{Schinke1993}. In case where there are many
overlapping levels due to multiple coupled potentials a such situation can
occur in which two very close oscillation energy levels can only be resolved
when there are more significant digits available than in the typical
\texttt{double} precision. The caveat being that a very small timestep is necessary
because it determines the resolution of the Fourier transform used on the autocorrelation function.
A work exploring this possiblity is currently in preparations to publish.}
\end{enumerate}}
%
%
%
\revStyle{Therefore,} I have implemented the algorithm presented here in high-precision.  In
\Section{sec:High:Precision} I show the performance benchmarks with additional
details.  See also my work~\cite{Kozicki2022a} for high-precision classical
dynamics.


\section{Implementation of TDSE for time-dependent Hamiltonian}
\label{chapter:Time:Dependent}

The C++ implementation of the iterative process in~\Eq{eqSemi62a} is shown in the \listing{listing4}
of the function \texttt{SemiGlobalODE}\hspace{0pt}\texttt{::propagateByDtSemiGlobal}~\cite{Kozicki2022b}.

First the simulation parameters (see~\tab{tab:paramsKM}) are assigned to local
variables.  A helper class \texttt{LocalData} in line 6 contains variables
local to this function as well as methods operating on them. It was introduced
here in order to improve clarity of the rest of the
\texttt{SemiGlobalODE}\hspace{0pt}\texttt{::propagateByDtSemiGlobal} by
splitting the calculation into several logical steps, described below.  A
matrix \texttt{Unew} (variable $\upsiBN_{new}$ in~\eq{eqSemi62a}) in line 9 is
created to store $M$ ($M$ is \texttt{Nt\_ts} in the code)
wavefunctions\footnote{Using plural \textit{wavefunction\textbf{s}} might be
confusing, so here's a clarification:
on each coupled electronic state there is a wavefunction evolving on the
electronic potential assigned to it. All of them together sit inside a C++
\texttt{std::vector} container.  When there is no confusion in the context I am
using \textit{wavefunction} to refer to all coupled
\textit{wavefunction\textbf{s}}, otherwise I emphasize in text whether I mean a
single coupled wavefunction or all wavefunctions. In here it's list of wavefunctions
one for each \revA{time point} in \Fig{fig:chebT}.}, one for each \revA{time point}
(\Fig{fig:chebT}). Additionally one extra $M+1^{\textrm{th}}$ wavefunction is
created for the purposes \revA{of estimation of the interpolation in time}.
The self convergent iteration process
(\eq{eqSemi62a}) spans whole $\Delta t$ time period divided into $M$ \revA{time points}
(\Fig{fig:chebT}). The guess wavefunction \texttt{d.Ulast} in line 12 (variable
$\upsiBN_{prev}$ in~\Eq{eqSemi62a}) takes value from the extrapolated
\texttt{Uguess} which was prepared at the end of the previous timestep (line~60
or 63)%
\footnote{When jump starting the calculations \texttt{Uguess} equals the
initial wavefunction (see \eq{eqSemi37}). The assignment to \texttt{Uguess} is
performed in the class constructor, hence it is not shown in the \Listing{listing4}.}.
In case when there was no previous iteration the $\upsiBN(t=0)$ is used for all
$M$ \revA{time point} wavefunctions, this causes the main convergence \texttt{while}
loop to be executed about 2 times more. The value $\upsiBN_{new}(\Delta t_0)$
at first \revA{time point} is the final value from previous timestep (line 13). The
$\vBN_j$ vectors (\eq{eqSemi47}) are created in line 14, their first component
is just the wavefunction $\upsiBN_0$ (\eq{eqSemi37}). In the main convergence
loop they are calculated in line~21. Remaining preparation lines concern
counting the number of iterations it took~\eq{eqSemi62a} to converge and
tracking and estimating the calculation error. The \texttt{tol+1} in line 16 is
to ensure that the first execution of the \texttt{while} loop always takes
place. In next executions of this loop the value from~\eq{eqSemiVII} is used
and compared against $\varepsilon$ tolerance.

%

\begin{figure*}
\begin{minipage}{\textwidth}
\begin{lstlisting}[numbers=left,firstline=2,lastline=69,label=listing4,
caption={The main time propagation loop for the time-dependent Hamiltonian},
language=c]
void SemiGlobalODE::propagateByDtSemiGlobal(const int& iteration)
{
	const int&  Nt_ts = cpar.M_Nt_ts; // number of interior Chebyshev points (M in article)
	const int&  Nfm   = cpar.K_Nfm;   // number of terms for function of matrix (K in article)
	const Real& tol   = cpar.tol;     // tolerance parameter 
	LocalData   d(*this, iteration);  // data with local variables e.g. Eq.15 and Eq.19
	// Unew is the calculated wavefunction. Uguess is prepared at the end of iteration
	// and used later. The extra Nt_ts + 1 column is used for estimating errors.
	Unew              = MatrixXcr::Zero(cpar.Nu, Nt_ts + 1);
	// The first guess for the iterative process, for the convergence of the u values.
	// Each column represents an interior time point in the time step:
	d.Ulast           = Uguess; // first guess for Eq.28
	Unew.col(0)       = d.Ulast.col(0);
	d.v_vecs.col(0)   = d.Ulast.col(0); // Eq.18 and 19
	niter             = 0;      // count iterations in Eq.28
	currentErrors     = EstimatedErrors(0, tol + 1); // error tracking

	// main iteration loop in Eq.28
	while (not d.converged(currentErrors, tol)) { // Eq.27
		d.calcSExtended(); // calculate Eq. 24, inhomogeneos term
		d.calcVVectors();  // calculate Eq. 18 and Eq. 19
		d.checkAllFinite(currentErrors); // check for divergence
		// The G_avg operator acting on v
		auto G_avg = [=](const VectorXcr& v) -> VectorXcr {
			return Gop(d.Ulast.col(cpar.tmidi), d.t(cpar.tmidi), v);
		};
		if (cpar.Arnoldi) { // use Arnoldi method
			// Create the orthogonalized Krylov space by the Arnoldi iteration procedure, Sect. 2.6
			const MatrixXcr Hessenberg = createKrop(G_avg, d.v_vecs.col(Nt_ts), Nfm, d.Upsilon);
			const VectorXcr eigval     = eigenValues(Hessenberg.block(0, 0, Nfm, Nfm));
			const Complex   avgp       = eigval.sum() / Real(Nfm);
			// sampled energy spectrum and extra point to estimate error
			d.samplingp.resize(eigval.rows() + 1);
			d.samplingp << eigval, avgp;
			d.capacity = get_capacity(eigval, avgp);
			// Obtain the expansion vectors for Newton approximation of f(G,t)v_vecs in Krylov space
			d.RvKr = getRv(Nfm, d.v_vecs, Nt_ts, Hessenberg, d.samplingp, d.capacity);
			// use Arnoldi to approximate new iteration of Eq.28
			Unew.block(0, 1, Unew.rows(), Unew.cols() - 1)
			    = Ufrom_vArnoldi(cpar.timeMts, tol, Nfm, d, currentErrors);
		} else { // Use a Chebyshev approximation for the function of matrix computation. Vcheb has:
			// Tn(G(Ulast(:,tmidi), t(tmidi)))*v_vecs(: ,Nt_ts + 1),  n = 0, 1, ..., Nfm-1
			// where Tn(z) are the Chebyshev polynomials, n'th vector is the (n+1)'th col of Vcheb.
			d.Vcheb = vchebMop(G_avg, d.v_vecs.col(Nt_ts), cheb->min_ev, cheb->max_ev, Nfm);
			Unew.block(0, 1, Unew.rows(), Unew.cols() - 1)
			    = Ufrom_vCheb(cpar.timeMts, cheb->Ccheb_f_ts, maxErrors, d, currentErrors);
		}
		d.estimateErrors(currentErrors);
		d.Ulast = Unew;
		niter   = niter + 1;
	}
	currentErrors.checkNiterIsEnough({ cpar, tol, iteration, niter, *this }); // Check if converged.
	maxErrors.updateMaxErrors(currentErrors); // update estimated maximum errors
	if (iteration == 0) niter0th = niter;
	allniter   = allniter + niter;
	fiSolution = Unew.col(Nt_ts - 1); // Finally! store the result wavefunction.
	// The new guess is an extrapolation of the solution within the previous time step:
	Uguess.col(0) = Unew.col(Nt_ts - 1); // result at last timestep is the first one in next iter
	if (cpar.Arnoldi) { // Arnoldi method
		Uguess.block(0, 1, Uguess.rows(), Nt_ts) // extrapolate for the next timestep
		        = Ufrom_vArnoldi(cpar.timeMnext, tol, Nfm, d, boost::none);
	} else { // Chebyshev method
		Uguess.block(0, 1, Uguess.rows(), Nt_ts) // extrapolate for the next timestep
		        = Ufrom_vCheb(cpar.timeMnext, cheb->Ccheb_f_next, maxErrors, d, boost::none);
	}
	if (there_is_ih) { sNext = d.s.col(Nt_ts - 1); }
	maxErrors.checkAll({ cpar, tol, iteration, -niter, *this }); // check esitmated errors
}
\end{lstlisting}
\end{minipage}\\[-8mm]
\end{figure*}

\begin{figure*}[!ht]
\begin{minipage}{\textwidth}
\begin{lstlisting}[numbers=left,firstline=2,lastline=29,label=listing5,
caption={The time-dependent Hamiltonian operator for coupled electronic states in C++},
language=c]
MultiVectorXcr
State::calc_Hpsi(const MultiVectorXcr& psi_0, const Real& t)
{
    MultiVectorXcr psi_ret = getZeroMultiVectorXcr(psi_0);
    // The kinetic operator, acting only on the diagonal
    for (int j = 0; j < levels; j++) {
        psi_ret[j] = Ekin_single(psi_0[j]);
    }
    // The potential operator matrix acting on the wavefunction
    for (int j = 0; j < levels; j++) {
        for (int k = 0; k < levels; k++) {
            // if present obtain the time dependent potential
            if (hasTimeDependence(j, k)) {
                psi_ret[j] = psi_ret[j].array()
                    + timeDependentPotential(j, k, t).array()
                        * psi_0[k].array();
            } else {
                psi_ret[j] = psi_ret[j].array()
                    + potentialMatrix[j][k].array()*psi_0[k].array();
            }
        }
    }
    return psi_ret;
}
\end{lstlisting}

$~$\\[-12mm]
\begin{lstlisting}[numbers=left,firstline=2,lastline=12,label=listing6,
caption={The C++ wrapper for handling arbitrary number of coupled electronic states},
language=c]
VectorXcr State::minus_i_Hpsi__MultiVectorFlattened(
    const VectorXcr& /*u*/, // wf from this step, e.g. Bose-Einstein
    const Real&      t,     // time
    const VectorXcr& v      // wavefunction wf to propagate
)
{
    MultiVectorXcr psi_work = zero(wf);
    decompres(v, psi_work);
    psi_work = calc_Hpsi(psi_work, t);
    return -Mathr::I * flatten(psi_work);
};
\end{lstlisting}
\end{minipage}\\[-8mm]
\end{figure*}

\begin{figure*}[!ht]
\begin{minipage}{\textwidth}
\begin{lstlisting}[numbers=left,firstline=2,lastline=10,label=listing3,caption={The kinetic energy operator using FFT implemented in C++},language=c]
VectorXcr State::Ekin_single(const VectorXcr& psi_0)
{
	VectorXcr psi_1 = VectorXcr::Zero(points);
	doFFT(psi_0, psi_1);                            // ψ₁=                  ℱ(ψ₀)
	psi_1 = (psi_1.array() * minusKSqared.array()); // ψ₁=               -k²ℱ(ψ₀)
	doIFFT(psi_1, psi_1);                           // ψ₁=           ℱ⁻¹(-k²ℱ(ψ₀)) = ∇²ψ₀
	psi_1 *= (-PhysConst::hbarSqr / (2 * mass));    // ψ₁=    (-ℏ²(∇²ψ₀)/(2m))
	return psi_1;
}
\end{lstlisting}
\end{minipage}\\[-8mm]
\end{figure*}

In the main \texttt{while} loop (line 19, \eq{eqSemi62a}), first the extended
inhomogeneous source term $\sBN_e(\upsiBN(t),t)$ from~\eq{eqSemi59} is
calculated in line 20 taking into account all of the time dependence of the
Hamiltonian. Next the $\vBN_j$ vectors are calculated for all $M$. The Newton
interpolation polynomial (\Section{sec:NewtonInterpolation}) is used and the
divided differences calculations are performed in the process. This is followed
by a check for numerical divergence (line 22).  Next a lambda function for the
$\GBN_{avg}$ (line~24) is created to be used in the calculation of the first
term $f_M(\GBN_{avg})$ in \eq{eqSemi62a}.  In line 27 code branching occurs
depending on the type of the calculation method.  The Arnoldi method is
described in this paper (\Section{sec:Arnoldi}), the Chebyshev method (line 41)
is discussed in detail in~\cite{Schaefer2017} and is also possible to use here
although it is less useful because of the need to provide the energy range of
the Hamiltonian.
In line 29 the
\revA{representation of $\GBN_{avg}$ in the orthogonalized Krylov space is stored in the}
Hessenberg matrix and its eigenvalues are found (line 30). This is the place in
the Arnoldi algorithm which finds the range of the energy spectrum of the
Hamiltonian and makes it possible to calculate using complex potential
and simplifies a lot the usage of this algorithm. One additional point in the
spectrum named \texttt{avgp} (line 31) is used in order to track the
calculation
error and compute the energy spectrum capacity~\cite{TalEzer1989} (line 35).
Next the expansion vectors for the Newton approximation in the reduced Krylov
space are calculated (line 37) and then all $M+1$ wavefunctions $\upsiBN_{new}$
are calculated (\eq{eqSemi62a}, line~39). It is this line that the conversion
between Krylov space and position representation is performed using the
transformation matrix $\Upsilon$ (c.f.~point (d) in \Section{sec:Arnoldi}).
This follows by estimating current convergence (line 48) \eq{eqSemiVII} and assigning
$\upsiBN_{prev}$ to $\upsiBN_{new}$.  As mentioned earlier the lines 41-47
perform the same calculation, but with Chebyshev approach~\cite{Schaefer2017},
notably the energy range \texttt{min\_ev} and \texttt{max\_ev} (used in line
44) has to be provided.

When the iterative process is complete a check is performed whether the used
number of iterations was enough for convergence (line 52) and maximum estimated
errors are stored (line 53).  Next, the total number of iterations is stored
(line 55) in order to track the computational cost. Next, the solution is
stored in a class variable \texttt{fiSolution} (line~56). Finally the
wavefunction is extrapolated for the next timestep $\Delta t$ (line~60 and 63)
and the estimated errors are checked and stored (line 67).

The time-dependent Hamiltonian $\Hbn(t)$ is called in lines 20 and 25, when invoking the
\texttt{calcSExtended} and \texttt{Gop} functions. It is presented in~\listing{listing5}.
The time
dependence is encoded in line~13 of~\listing{listing5}, also it deals with coupled electronic states (\Section{sec:Coupled:Implementation}). The time-dependent potential is used in line~15 when
acting on the wavefunction in line~16.  The time-dependent Hamiltonian uses function
\texttt{Ekin\_single} (\Listing{listing3}) in line~7 (\Listing{listing5}) to calculate the kinetic energy operator although if necessary
it also could become time-dependent
with only a small change in the code.

\revB{It shall be noted here that the Listings~\ref{listing5}, \ref{listing6} and \ref{listing3}
are example implementations of a complete Hamiltonian together with coupled electronic
levels and a custom kinetic energy operator, which uses FFT. The \textit{semi-global}
algorithm in \Listing{listing4} can be supplied by the user with an
entirely different set of these functions
implementing a different Hamiltonian and kinetic energy operator. The possibilities include
(1) curvilinear coordinates with more degrees of freedom than the single degree of freedom
used here, e.g.~Jacobi coordinates (2) a grid with varying distance between the points.
The only requirement being that the implemented Hamiltonian correctly ,,understands''
the wavefunction and deals with it. The \textit{semi-global} algorithm does not need
to know the details how the wavefunction is dealt with by the Hamiltonian as long as
it is converted to \texttt{VectorXcr} (,,flattened'', see Sections
\ref{sec:multiLevel},
\ref{sec:Coupled:Implementation}
and the
file \texttt{README.pdf} in the accompanying source code package)}.

In \revB{the presented numerical implementation} the numerical damping absorbing
boundary conditions \revB{from} \Section{sec:dampingComplex} are
encoded inside the complex potential (lines 15 and 19~in~\listing{listing5}) during the
calculations.


\subsection{Coupled time-dependent Schrödinger equations}
\label{sec:Coupled:Implementation}

In \Section{sec:multiLevel} I explained how it is possible to adapt this
algorithm to arbitrary amount of coupled electronic states by storing all
levels inside a single table.

To deal with each \texttt{k}${}^{\rm{th}}$ level the \texttt{Ekin\_single}
loops over all \texttt{levels} (\listing{listing5} line~6). Similarly when
dealing with the off diagonal elements of \eq{eqSchTDHCoupled} the loop
on \texttt{levels} is done in lines~10 and 11.  The function
\texttt{calc\_Hpsi} in \listing{listing5} as the wavefunction argument takes
the \texttt{MultiVectorXcr} which contains all wavefunctions as separate
elements of \texttt{std::vector}\footnote{\label{ft:WFtype}To be precise in C++
the following types are defined:\\\texttt{using VectorXcr =
Eigen::Matrix<Complex, Eigen::Dynamic, 1>;}\\ \texttt{using MultiVectorXcr =
std::vector<VectorXcr>;}}. But the \textit{semi-global} algorithm deals with a single
,,flattened'' \texttt{VectorXcr}. Hence a conversion discussed in
\Section{sec:multiLevel} has to be performed.

This conversion is shown in~\listing{listing6}.  The
\texttt{minus\_i\_Hpsi\_}\hspace{0pt}\texttt{\_MultiVectorFlattened} is the
function which is provided to the \textit{semi-global} algorithm as the function
$\GBN(\upsiBN(t),t)$ (\eq{eqSchTDH3}, \listing{listing4} line~24).  It is
called with wavefunction stored inside a single argument \texttt{VectorXcr}.
This data is decompressed in line~8 (\listing{listing6}), then the
time-dependent Hamiltonian \texttt{calc\_Hpsi} is called on it
(\listing{listing5}) and then the data is flattened again into a single
\texttt{VectorXcr} and multiplied by negative imaginary unit (\eq{eqSchTDH3},
line~10~in \listing{listing6}) (atomic units are used here). The \textit{semi-global}
algorithm can also work when the wavefunction from present timestep is used
(e.g. a Bose-Einstein condensate trap) and provides this argument for the
Hamiltonian in line~2 of \listing{listing6}.  The wavefunction from present
timestep is the \texttt{d.Ulast.col(cpar.tmidi)} (line~25~in
\listing{listing4}) where \texttt{cpar.tmidi} is the time coordinate of the
averaged Hamiltonian $\GBN_{avg}$\footnote{In \listing{listing6} the variable
\texttt{/*u*/} is commented out because here I am not dealing
with the Bose-Einstein condensate. It is confirmed to work by comparing with
examples provided in~\cite{Schaefer2017}.}.

\clearpage
\section{Validation of the C++ code for semi-global algorithm}
\label{sec:validation}

To validate the implementation of the \textit{semi-global} time propagator I have
reproduced the results both from~\cite{Schaefer2017} and
from \cite{Tully1990,Coker1993}%
\footnote{Additionally I have compared the simulations of a two level NaRb
system
with the WavePacket
software~\cite{Schmidt2017,Schmidt2018,Schmidt2019}.
}.
The following tests are 
provided in the accompanying C++ code:

{\renewcommand{\labelenumi}{(\alph{enumi})}
\begin{enumerate}
\item Atom in an intense laser field,~\Section{sec:validate:atom} (accompanying file \texttt{test\_atom\_laser\_ABC.cpp}).

\item Single avoided crossing,~\Section{sec:singleCross} (accompanying file \texttt{test\_single\_dual\_crossing.cpp}).

\item Dual avoided crossing,~\Section{sec:dualCross} (accompanying file \texttt{test\_single\_dual\_crossing.cpp}).

\item Gaussian packet in a forced harmonic oscillator, supplementary materials
of~\cite{Schaefer2017} (found in the accompanying file \texttt{test\_source\_term.cpp}).

\item Forced harmonic oscillator with an arbitrary inhomogeneous source term, supplementary
materials of~\cite{Schaefer2017} (accompanying file \texttt{test\_source\_term.cpp}).

\end{enumerate}}

The tests with a Gaussian packet in a forced
harmonic oscillator and an oscillator with an inhomogeneous source term are
example simulations found in the supplementary materials of~\cite{Schaefer2017}
and I include them in the C++ code, but do not discuss them here.
Suffice to say that I have reproduced them exactly.

\subsection{Validation of time-dependent Hamiltonian using an atom in an intense laser field}
\label{sec:validate:atom}

Here I present the reproduction of results of a model atom in an
intense laser field which was presented in~\cite{Schaefer2017}. I shall note
that these are not new results, since validation of an algorithm has to be run
on an example for which the results are already known and verified. In this
case I~am comparing the electronic wavefunction of an atom in a laser field
after evolution for $1000$~a.u.~(a time of about 24.2~fs), with the reference
result found in the supplementary materials of~\cite{Schaefer2017}. In the
calculations I have used the following parameters (\tab{tab:paramsKM}): $K=9$,
$M=9$, $\varepsilon = 2\times10^{-16}$ (for \texttt{double} precision%
\footnote{for \texttt{long double} $\varepsilon = 1\times10^{-19}$ and
for \texttt{float128} $\varepsilon = 2\times10^{-34}$.}) and $\Delta t=0.025$~a.u.

In this test the central potential is represented by a simplified Coulomb
potential without the singularity (hence it is a model one dimensional
atom):\\[-5mm]

\begin{equation}
\label{eqAtom}
V_{atom}(x)=1-\frac{1}{\sqrt{x^2+1}}.
\end{equation}

\begin{figure}
    \centering
    \includegraphics[height=\columnwidth,angle=270]{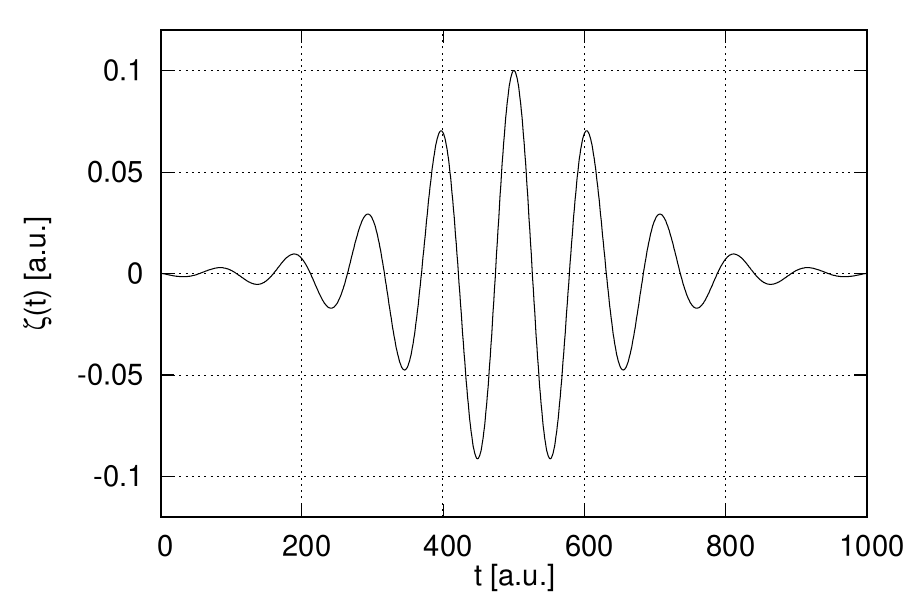}
    \caption{The electric field of the laser impulse~\cite{Schaefer2017}.}
    \label{fig:AtomLaserE}
\end{figure}

This simple model is for example used in the context of intense laser atomic
physics.
The electric field of the laser impulse used is following
(\fig{fig:AtomLaserE}):

\begin{equation}
\label{eqEfield}
    \zeta(t)=0.1{\textrm{sech}^2\left(\frac{t-500}{170}\right)}\cos(0.06(t-500)).
\end{equation}

This laser impulse with $\omega=0.06$~a.u.~is similar to the wavelength of a
Titanium-Sapphire laser which is $\lambda=760$~nm. The sech$^2=1/\cosh^2$
envelope is similar to the actual envelope found in the laser pulses. The
maximum amplitude ${\zeta_{max}=0.1}$~a.u. represents the intensity of about
$I_{max}=3.52\times 10^{14}$~W/cm$^2$. This term, together with the dipole
approximation $x\,\zeta(t)$ is added to the Hamiltonian which reads:

\begin{equation}
\label{eqAtomHamilt}
\Hbf(t)=\frac{-\hbar^2}{\revB{2}} \nabla^2 + 1-\frac{1}{\sqrt{x^2+1}}-x \zeta(t).
\end{equation}

\begin{figure*}
\begin{center}
\setlength{\unitlength}{1.75mm}
\begin{picture}(90,48)
\put(4.3,-6){\begin{picture}(90,55)
\put(2,6){\includegraphics[width=157.5mm]{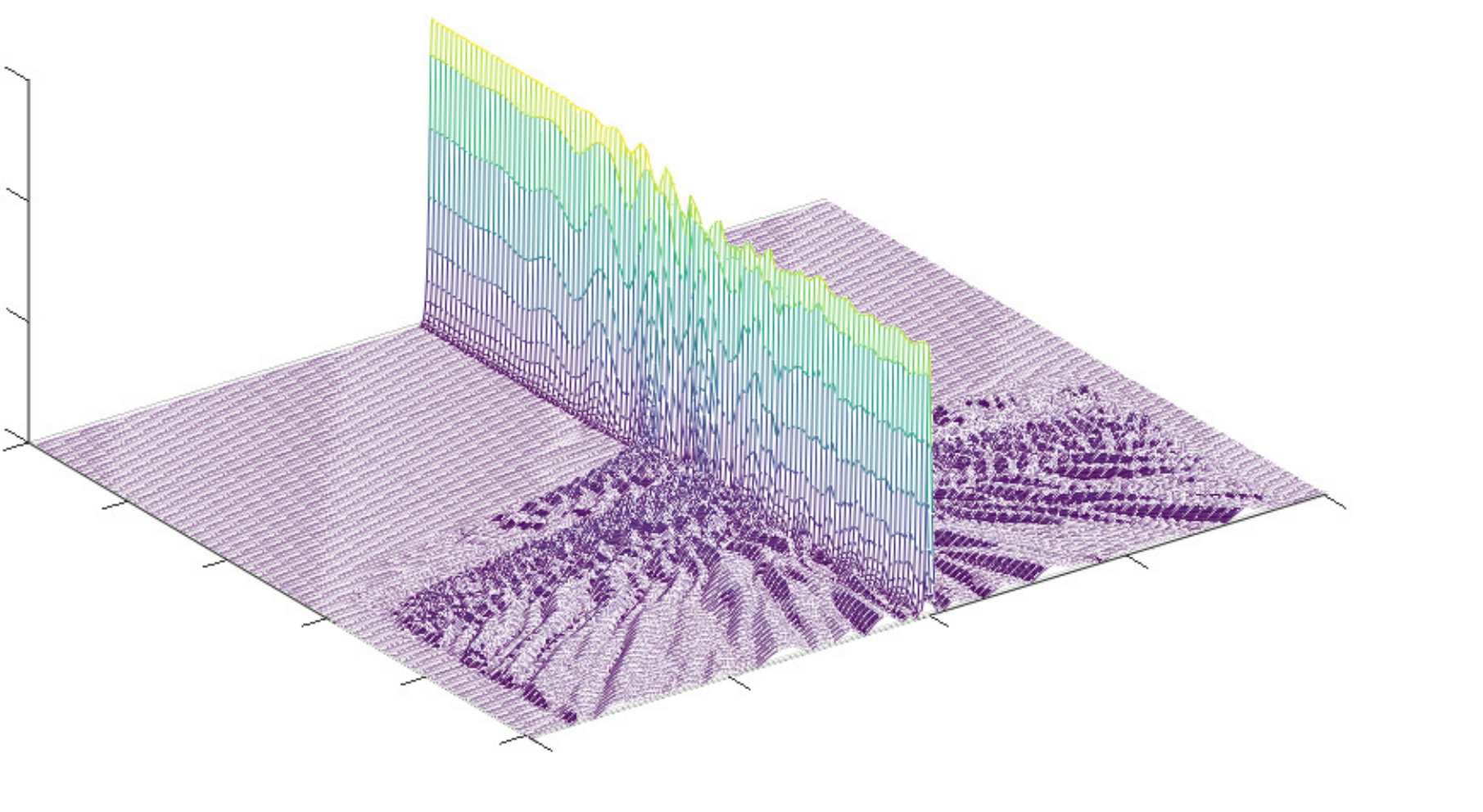}}
\put(0.5,27){\small{0}}
\put(5,23){\small{200}}
\put(11,19.6){\small{400}}
\put(17,16.2){\small{600}}
\put(23,12.8){\small{800}}
\put(28,9.4){\small{1000}}
\put(7,16){{\small{$t[a.u.]$}}}
\put(0.5,29){\small{0}}
\put(-1,36){\small{0.2}}
\put(-1,43.5){\small{0.4}}
\put(-1,51){\small{0.6}}
\put(-5,38){\rotatebox{90}{\small{$|\psi(R)|^2$}}}
\put(35,8){\small{-240}}
\put(47,11.65){\small{-120}}
\put(60,15.3){\small{0}}
\put(72,18.95){\small{120}}
\put(84,22.6){\small{240}}
\put(58,12){{\small{$R[a_0]$}}}
\end{picture}}
\end{picture}
\end{center}
\caption[Model atom in an intense laser field, the time evolution of $|\psi(R)|^2$]{Model atom in an intense laser field, the time evolution of $|\psi(R)|^2$ during 1000~a.u. The wavefunction at 1000~a.u. agrees with reference solution in~\cite{Schaefer2017} with error $<$ $8\times10^{-15}$.}
\label{fig:AtomLaserWF}
\end{figure*}
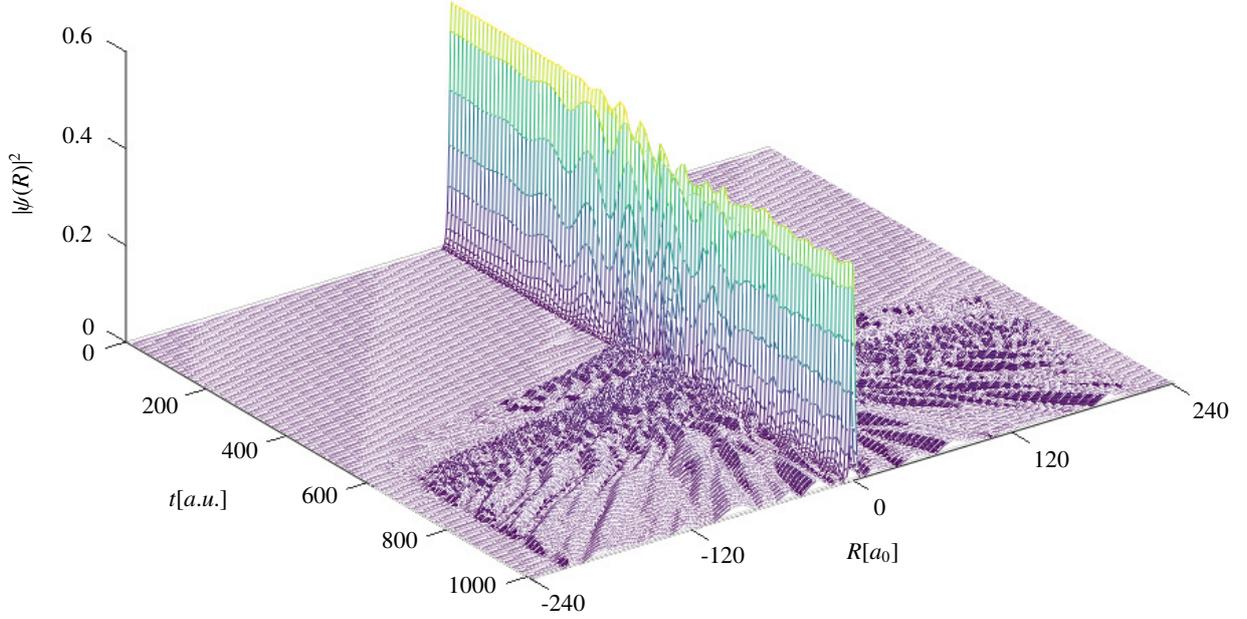

\noindent
For the purposes of numerical
simulation this Hamiltonian is modified by adding complex absorbing boundary
conditions (\Section{sec:dampingComplex}).  Their effect can be seen on the
left and right edges of \fig{fig:AtomLaserWF} where the wavefunction is
diminished.  The initial wavefunction is the ground state of the model atom.

\Fig{fig:AtomLaserWF} shows the time evolution of the wavefunction over the
time of 1000~a.u. The maximum intensity of the laser impulse is centered on
500~a.u.  (\fig{fig:AtomLaserE}) and it can be seen on \fig{fig:AtomLaserWF}
that this is when the wavefunction starts to undergo a rapid change and some
parts of the wavefunction start the process of dissociation. When the
calculations reach the 1000~a.u.~point in time I compare the results with the
reference solution found in the supplementary materials in~\cite{Schaefer2017}
and the maximum difference at a grid point is smaller than~$8\times10^{-15}$, well within the
range of the numerical ULP error\footnote{see
footnote\textsuperscript{\ref{foot:ULP}} on page \pageref{foot:ULP} for
details.}
. It means
that my implementation of the \textit{semi-global} algorithm completely reproduces the
reference results.

\subsection{Validation of coupled Schrödinger equations using standard benchmarks}
\label{sec:validate1}


To validate \revStyle{the} present implementation, in this section, I will reproduce the
solution of two problems for a system of coupled Schrödinger equations
featuring nonadiabatic quantum dynamics.  They were originally introduced
in~\cite{Tully1990} and afterward carefully analyzed in~\cite{Coker1993}.
\revStyle{They} are considered to be \revStyle{the} standard benchmark for coupled systems.
In these problems the diagonal of the diabatic potential energy surfaces
$V_{11}(R)$ and $V_{22}(R)$ undergoes:


{\renewcommand{\labelenumi}{(\alph{enumi})}
\begin{enumerate}
\item a single crossing as in \fig{fig:potMatrix1}a and
\item a dual crossing   as in \fig{fig:potMatrix2}a.
\end{enumerate}}

\noindent
In the crossing region there is a strong coupling $V_{12}(R)$ between the two
levels.  In adiabatic representation the potential energy surfaces $E_1(R)$ and
$E_2(R)$ undergo respectively a single (\Section{sec:singleCross} and
\fig{fig:potMatrix1}b) and dual (\Section{sec:dualCross} and
\fig{fig:potMatrix2}b) avoided crossing.  The nonadiabatic coupling matrix
elements $D_{12}(R)$ between the two levels are relatively large
(\fig{fig:potMatrix1}b and \fig{fig:potMatrix2}b).

The calculations are performed diabatically because the coupling elements are
smaller and the Hamiltonian assumes a simpler
form~\cite{Baer2006,Tannor2007}\footnote{See Eqs.~2.12 and
2.100~in~\cite{Baer2006} for adiabatic TDSE:
$i\hbar\frac{\partial\psi}{\partial t}=-\frac{\hbar^2}{2m}(\nabla+\tau)^2\psi$
(where $\tau$ is the nonadiabatic coupling matrix) and Eqs.~2.22 and 2.115 for
diabatic TDSE: $i\hbar\frac{\partial\chi}{\partial
t}=\left(-\frac{\hbar^2}{2m}\nabla^2+V\right)\chi$; it can be seen that the
position of $\tau$ is rather unfortunate in the adiabatic representation.}.
To gain more insight from the wave packet dynamics on each of these levels the
results are presented here in adiabatic representation.  The two levels
$E_1(R)$ and $E_2(R)$ do not cross and the evolution of wavefunction
probability distributions (\figs{fig:repr10i12} and \ref{fig:reprNNi14}) on the
lower and the higher electronic surface is easier to interpret.  The
transmission and reflection probabilities (\figs{fig:repr11i13} and
\ref{fig:repr15i16}) on $E_1(R)$ and $E_2(R)$ are calculated as integrals (on a
discrete grid) of the single coupled wavefunction in the nuclear coordinate
range satisfying $R>0$ and $R<0$ respectively.  To obtain these results the
wavefunctions are converted  from diabatic representation to adiabatic
representation by
first performing the diagonalization of the potential matrix $V(R)$ for each
value of $R$. The obtained eigenvalues are the adiabatic potential energy
surfaces $E_1(R)$ and $E_2(R)$~\cite{Coker1993}. Next, the matrix of two
eigenvectors $\phi_{1}(R)$ and $\phi_{2}(R)$ (following the method presented in \cite{Coker1993}):\\[-7mm]

\begin{equation}
\label{eqCoker65}
P(R)=\left[\begin{matrix}P_{11}(R) & P_{12}(R) \\ P_{21}(R) & P_{22}(R)\end{matrix}\right]=%
\left[\,\,\phi_{1}(R)\,\,\,\,\, \phi_{2}(R)\,\,\right],
\end{equation}

\noindent
is used to compute the nonadiabatic coupling matrix elements $D_{12}(R)$ with:

\begin{equation}
\label{eqCoker66}
D_{12}(R)=\phi_{1}^{*} \cdot \frac{\partial}{\partial R} \phi_{2}=P_{11}^{*} \frac{d P_{12}}{d R}+P_{21}^{*} \frac{d P_{22}}{d R}.
\end{equation}

\noindent
Above the electronic integrals become a dot product or a sum in a two state
basis. Finally to obtain the adiabatic wavefunction $\psi(R,t)$ from the
diabatic one, $\chi(R,t)$, the following transformation is used:

\begin{equation}
\label{eqCoker69}
\psi(R,t)=P(R)\chi(R,t)
\end{equation}

\noindent
or more specifically, since in these two examples we are dealing with a two
level system\footnote{Alternative method of obtaining the transformation matrix
$P(R)$ is to use the Eq.~3.45 from~\cite{Baer2006}:
${\beta(R)=-\frac{1}{2}\tan^{-1}\left(\frac{V_{12}(R)}{V_{11}(R)}\right)}$ and
then use the rotation matrix from in Eq.~3.44~\cite{Baer2006}.}:

\begin{equation}
\label{eqCoker69more}
\left[\begin{matrix}\psi_{1}(R,t) \\ \psi_{2}(R,t)\end{matrix}\right]=%
\left[\begin{matrix}P_{11}(R) & P_{12}(R) \\ P_{21}(R) & P_{22}(R)\end{matrix}\right]%
\left[\begin{matrix}\chi_{1}(R, t) \\ \chi_{2}(R, t)\end{matrix}\right].
\end{equation}


\bigskip

The simulations begin with the single wavefunction placed on the lower energy
surface $E_1(R)$ assuming shape of a
Gaussian wavepacket\footnote{To be precise
in the code the following formula is used:
$\chi_{Gauss}(R)=\frac{\exp \left(-\frac{m (R-R_0)^2+i a^2 k_0 (k_0 \hbar  (t-t_0)-2 m (R-R_0))}{2 a^2 m+2 i \hbar (t-t_0)}\right)}{\sqrt{\sqrt{\pi } \left(a+\frac{i \hbar  (t-t_0)}{a m}\right)}}$,
as it is the solution of a free propagating Gaussian wave packet, in
\eq{eqGaussian} it was assumed that $t_0 = 0$.}:

\begin{equation}
\label{eqGaussian}
\chi_{Gauss}(R)=
{\pi^{-\frac{1}{4}}a^{-\frac{1}{2}}}
{e^{{}^{-\frac{(R-R_0)^2}{2 a^2}-{i k_0 (R-R_0)}}}}.
\end{equation}

\noindent
The initial values of parameters assumed in each simulation are listed in
\tab{tab:gaussParams}. They were chosen so as to best represent the undergoing
evolution of the coupled system dynamics and to completely reproduce the
results in~\cite{Tully1990,Coker1993}.


\def\SP#1{{\hspace{7.48mm}}\textrm{#1}\hspace{7.38mm}}

\begin{table*}[ht]
\begin{center}
\caption[Gauss wavepacket parameters used to reproduce the standard benchmark]{Gauss wavepacket parameters used to reproduce the standard benchmark~\cite{Tully1990,Coker1993}.}%
\label{tab:gaussParams}
\begin{tabular*}{0.81\textwidth}{l*{6}{S}}
\toprule
\multicolumn{2}{l}{parameter [a.u.]}& \multicolumn{2}{c}{single crossing} & \multicolumn{2}{c}{dual crossing} \\
\cmidrule(lr){3-6}
                         &         & \SP{high $k_0$}   &\SP{low $k_0$}   & \SP{high $k_0$} & \SP{low $k_0$}  \\
\toprule
mass                     & $m$     & 2000              & 2000            & 2000            & 2000            \\
wavenumber               & $k_{0}$ & 15                & 8.5             & 52              & 30              \\
packet width             & $a$     & 0.75              & 0.8             & 0.7             & 0.7             \\
start position           & $R_{0}$ & -4                & -4.15           & -8              & -8              \\
simulation time          & $T$     & 1200              & 4000            & 900             & 1500            \\
\bottomrule
\end{tabular*}

\smallskip
\caption{\revA{The parameters used in precision and performance tests of \textit{semi-global} method (see \Tab{tab:paramsKM}).}}
\label{tab:paramsPP}
\begin{tabular*}{0.81\textwidth}{*{5}{l}}
\toprule
Precision               & $K$ & $M$ & $\varepsilon$ ~~(equal to ULP size of given precision)                        & $\Delta t$ \\
\midrule
\texttt{double}         & 15  &  3  & $2.220446049250313\times10^{-16}$                   & 1 a.u. \\
\texttt{long double}    & 18  &  3  & $1.084202172485504434\times10^{-19}$                & 1 a.u. \\
\texttt{boost float128} & 31  &  3  & $1.925929944387235853055977942584927\times10^{-34}$ & 1 a.u. \\
\bottomrule
\end{tabular*}
\smallskip
\caption{\revA{Precision test of results from global Chebyshev propagator~\cite{Kosloff1984} compared against itself, but with higher computation precision. Error is given in terms of ULP (see footnote\textsuperscript{\ref{foot:ULP}}), where the size of ULP for each row is given in the last column (its precise value is in~\Tab{tab:paramsPP}). There is no row for \texttt{boost float128} because there is no result with higher precision against which it can be compared, hence it is used only as reference against which lower precision results are compared.}}
\label{tab:precChebySelf}
%
\begin{tabular}{l l r r l}
\toprule
Benchmark type & Precision & \texttt{long double} & \texttt{boost float128} & ULP size \\
\midrule
single crossing & \texttt{double}         & 143                  &  141   & $2.2\times10^{-16}$    \\
high $k_0$      & \texttt{long double}    & ---                  &  3824  & $1.1\times10^{-19}$ \\
\midrule
single crossing & \texttt{double}         & 127                  &  127   & $2.2\times10^{-16}$    \\
low $k_0$       & \texttt{long double}    & ---                  &  1026  & $1.1\times10^{-19}$ \\
\midrule
dual crossing & \texttt{double}         & 132                  &  124   & $2.2\times10^{-16}$    \\
high $k_0$    & \texttt{long double}    & ---                  &  15671 & $1.1\times10^{-19}$ \\
\midrule
dual crossing & \texttt{double}         & 733                  &  743   & $2.2\times10^{-16}$    \\
low $k_0$     & \texttt{long double}    & ---                  &  20242 & $1.1\times10^{-19}$ \\
\bottomrule
\end{tabular}

\end{center}
\end{table*}

\undef\SP

\subsubsection{Single avoided crossing}
\label{sec:singleCross}

\begin{figure}[ht]
    \centering
    \vtop{\includegraphics[width=\columnwidth]{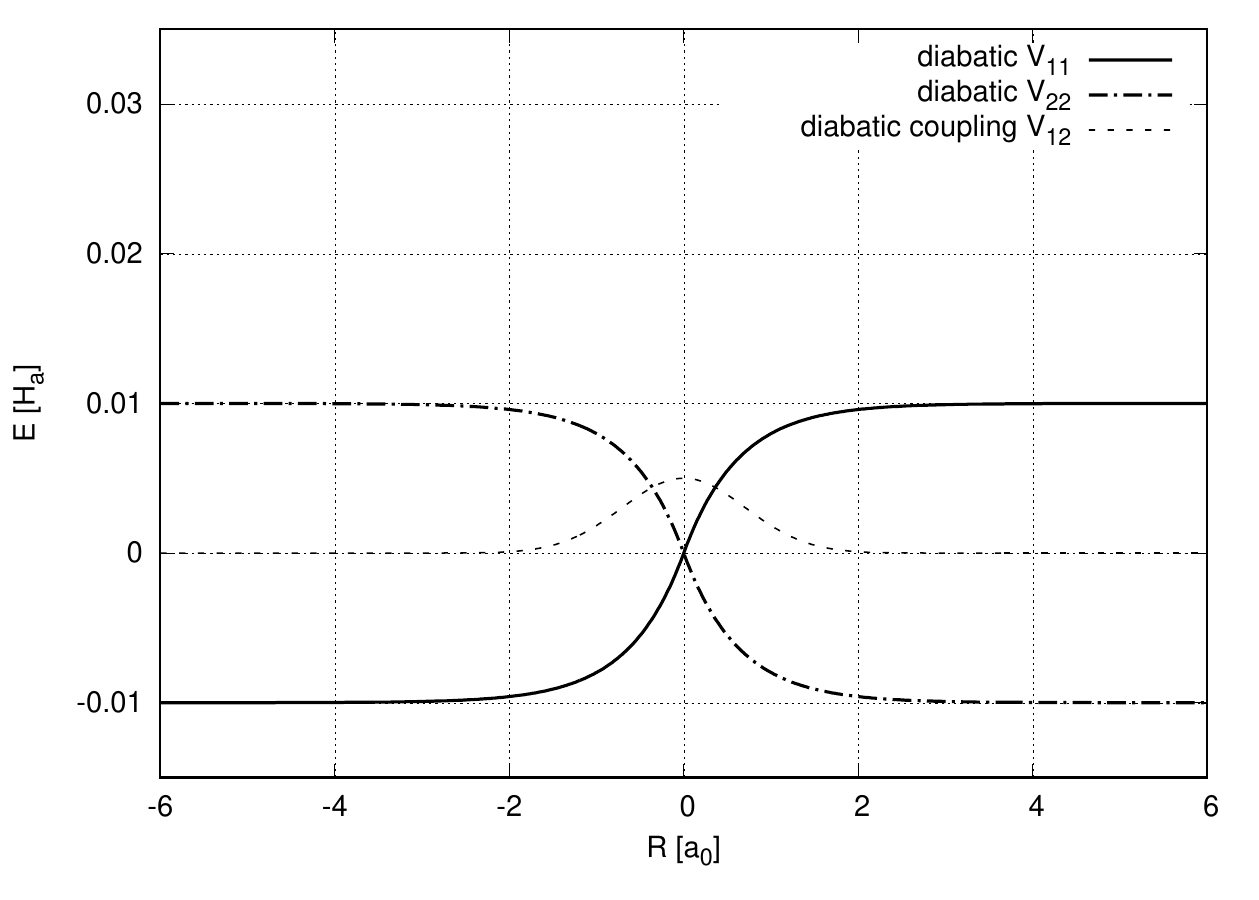}\\\hspace{7mm}(a)}%
    \vtop{\includegraphics[width=\columnwidth]{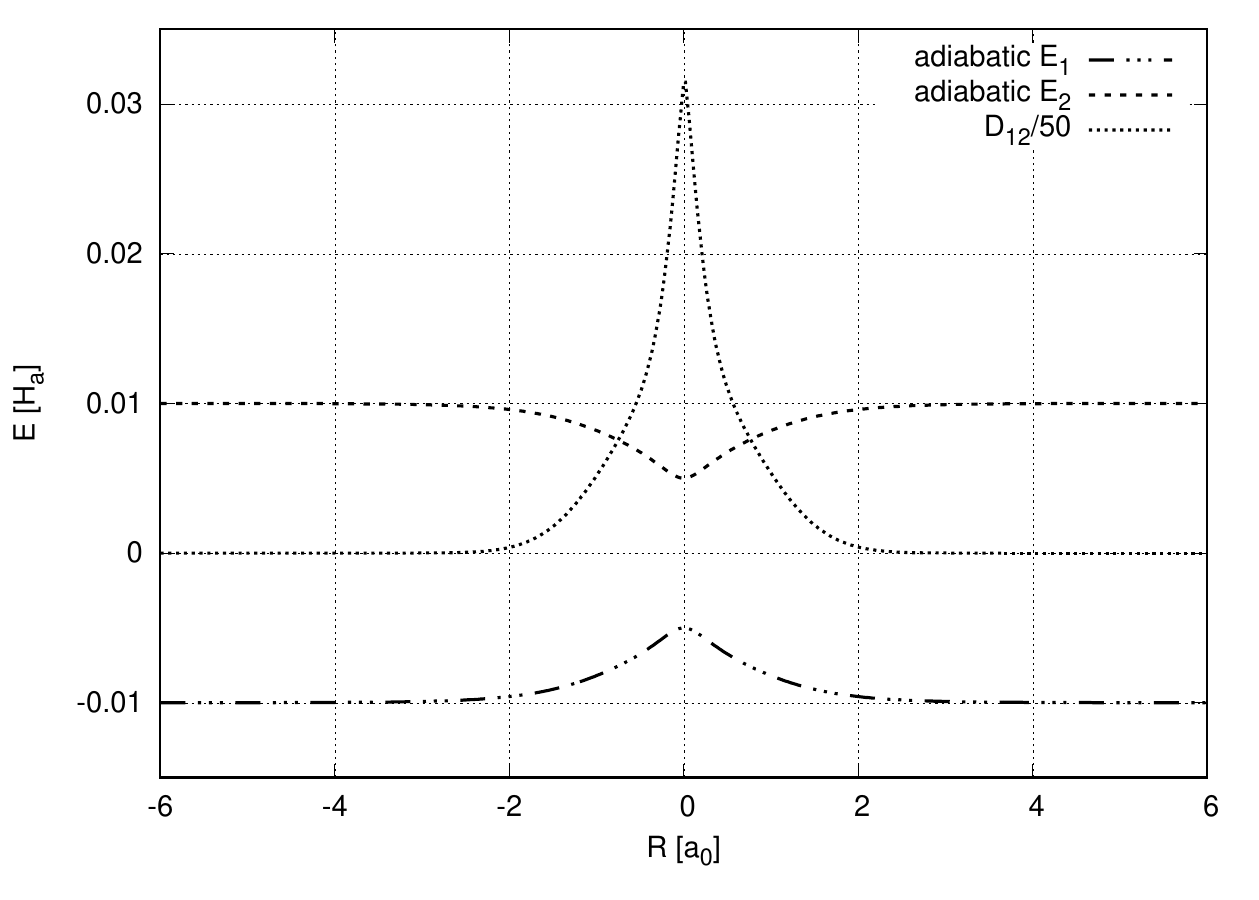}\\\hspace{7mm}(b)}%
    \caption[Model surfaces potential matrix of the simple avoided crossing example]{Model surfaces potential matrix of the simple avoided crossing example;
    (a)~diabatic representation;
    (b)~adiabatic representation, $D_{12}$ is the nonadiabatic coupling element (it is drawn as divided by 50 because the coupling is large).
    }
    \label{fig:potMatrix1}
\end{figure}

In the first standard benchmark~\cite{Tully1990,Coker1993} the potential matrix
elements in diabatic representation $V_{ij}(R)$ are defined as follows
(\fig{fig:potMatrix1}a):

\begin{equation}
\label{eqSimpleCrossing}
\begin{aligned}
V_{11}(R)&=\left\{
    \begin{array}{ll}
    A\left(1-e^{-B R}\right) & \textrm{~~for~~~~} R\geq 0 \\[1mm]
    -A\left(1-e^{B R}\right) & \textrm{~~for~~~~} R<0
    \end{array}
    \right.\\
V_{22}(R)&=-V_{11}(R) \\
V_{12}(R)&=V_{21}(R)=C e^{-D R^{2}}
\end{aligned}
\end{equation}

\noindent
with the parameters assuming values:
$A= 0.01$,
$B= 1.6$,
$C= 0.005$ and
$D= 1.0$.
In this example the diabatic surfaces cross at nuclear coordinate $R=0$ and a
Gaussian off-diagonal potential is assumed centered at this point. The
adiabatic surfaces $E_1(R)$ and $E_2(R)$ (\fig{fig:potMatrix1}b) repel each
other in the strong-coupling region and a large nonadiabatic coupling element
$D_{12}$ (\eq{eqCoker66}) appears at the avoided crossing.



\begin{figure}[ht]
    \centering
    \vtop{\includegraphics[width=\columnwidth]{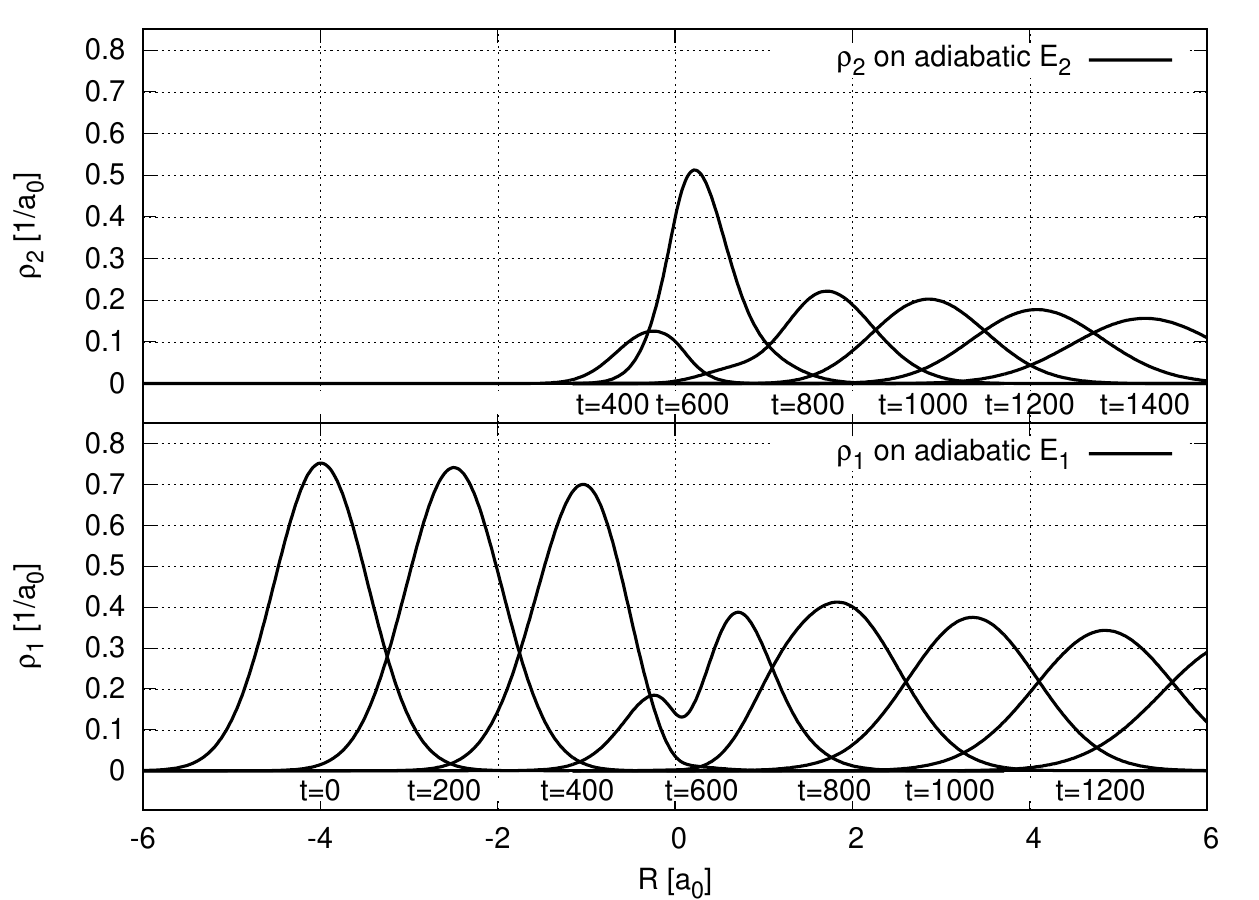}\\\hspace{7mm}(a)}%
    \vtop{\includegraphics[width=\columnwidth]{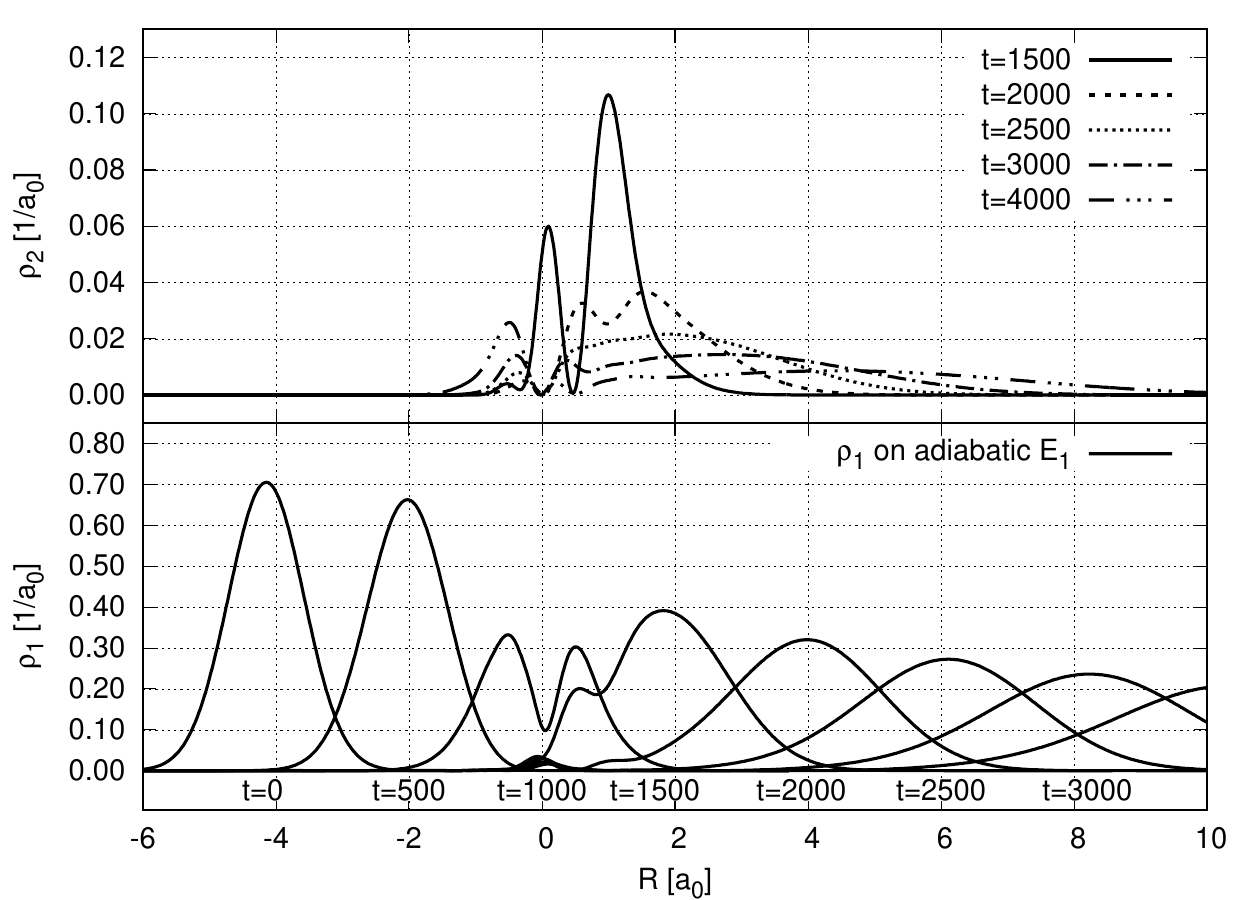}\\\hspace{7mm}(b)}%
    \caption[Evolution of adiabatic probability distributions in simple avoided crossing]{Time evolution of probability distributions on adiabatic energy surfaces, simple avoided crossing;
    (a)~high momentum Gaussian wavepacket $k=15$~a.u.;
    (b)~low momentum Gaussian wavepacket $k=8.5$~a.u.
    }
    \label{fig:repr10i12}
\end{figure}

\begin{figure}[ht]
    \centering
    \vtop{\includegraphics[width=\columnwidth]{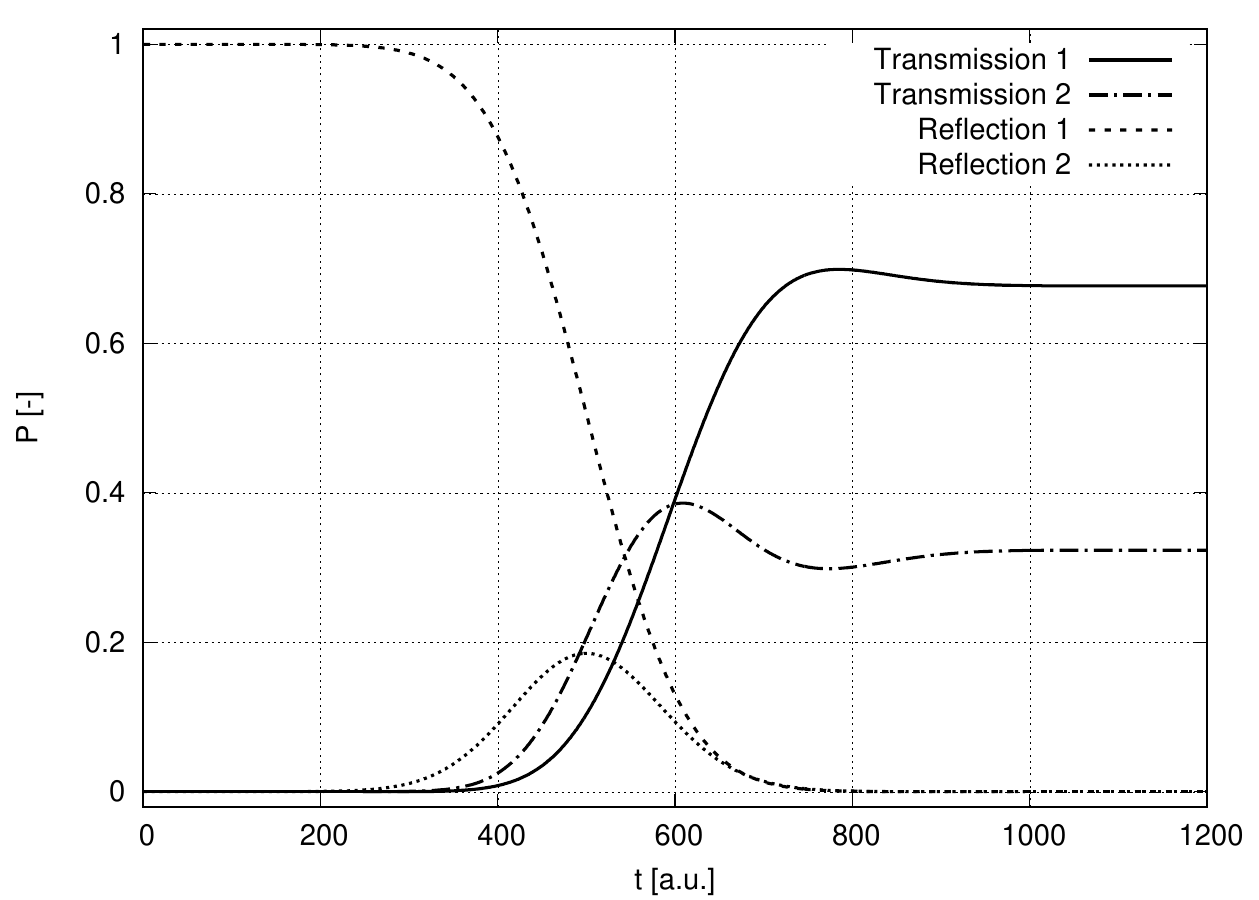}\\\hspace{7mm}(a)}%
    \vtop{\includegraphics[width=\columnwidth]{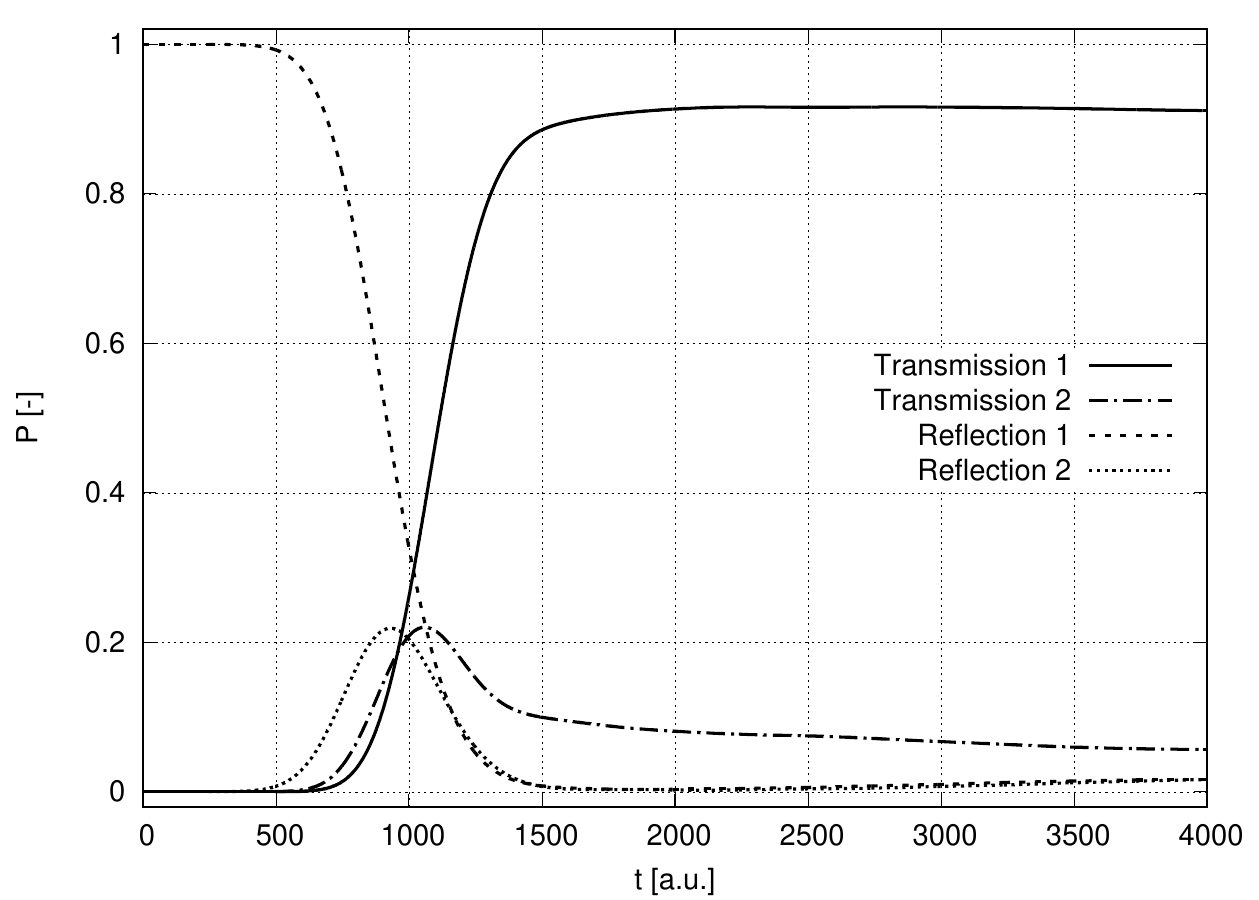}\\\hspace{7mm}(b)}%
    \caption[Transmission and reflection probabilities in simple avoided crossing]{Transmission and reflection probabilities as function of time in simple avoided crossing;
    (a)~high momentum Gaussian wavepacket $k=15$~a.u.;
    (b)~low momentum Gaussian wavepacket $k=8.5$~a.u.
    }
    \label{fig:repr11i13}
\end{figure}

\Fig{fig:repr10i12} shows the time evolution of the probability distributions
(squared amplitude of the wavefunction $\left|\psi(R,t)\right|^2$) on each of
the adiabatic electronic surfaces. And \fig{fig:repr11i13} shows the
transmission and reflection probabilities (integrals for $R>0$ and $R<0$)
evolving over time.

In the high momentum case (\fig{fig:repr10i12}a) the packet has enough energy
to put about 32\% 
of the population on the upper energy surface. Entering higher level $E_2(R)$
caused the wave packet to lose energy, it has smaller momentum and is
propagating slower 
as can be seen by the time labels put beneath the center of each packet in the
\fig{fig:repr10i12}a.  \Fig{fig:repr11i13}a shows the transmission and
reflection probabilities for high momentum case. It can be seen that whole
packet passes through the crossing point and reflection vanishes over time. The
final population is about twice higher on the lower electronic surface than on
the upper one.

In the low momentum case (\fig{fig:repr10i12}b) the packet doesn't have enough
energy to populate the higher energy surface. Note the vertical scale on the
upper level $\rho_2$ in \fig{fig:repr10i12}b. After going up, the packet almost
does not move forward and instead it is leaking back to the lower level in both
directions. It can be seen in \fig{fig:repr11i13}b that the reflection weakly
increases over time while transmission on both electronic surfaces slowly
decreases. Most (91\%) of the final population resides on the lower surface.

The obtained results are in very good agreement
with~\cite{Tully1990,Coker1993}.  Please note that I obtained these results
using a different, arguably more precise, time
integration algorithm than the one used in~\cite{Tully1990,Coker1993}.
\revA{Also see \Section{sec:precision:performance} for precision and
performance comparison of these calculations against the
global Chebyshev propagator~\cite{Kosloff1984,Kosloff1997}}.

\subsubsection{Dual avoided crossing}
\label{sec:dualCross}

\begin{figure}[ht]
    \centering
    \vtop{\includegraphics[width=\columnwidth]{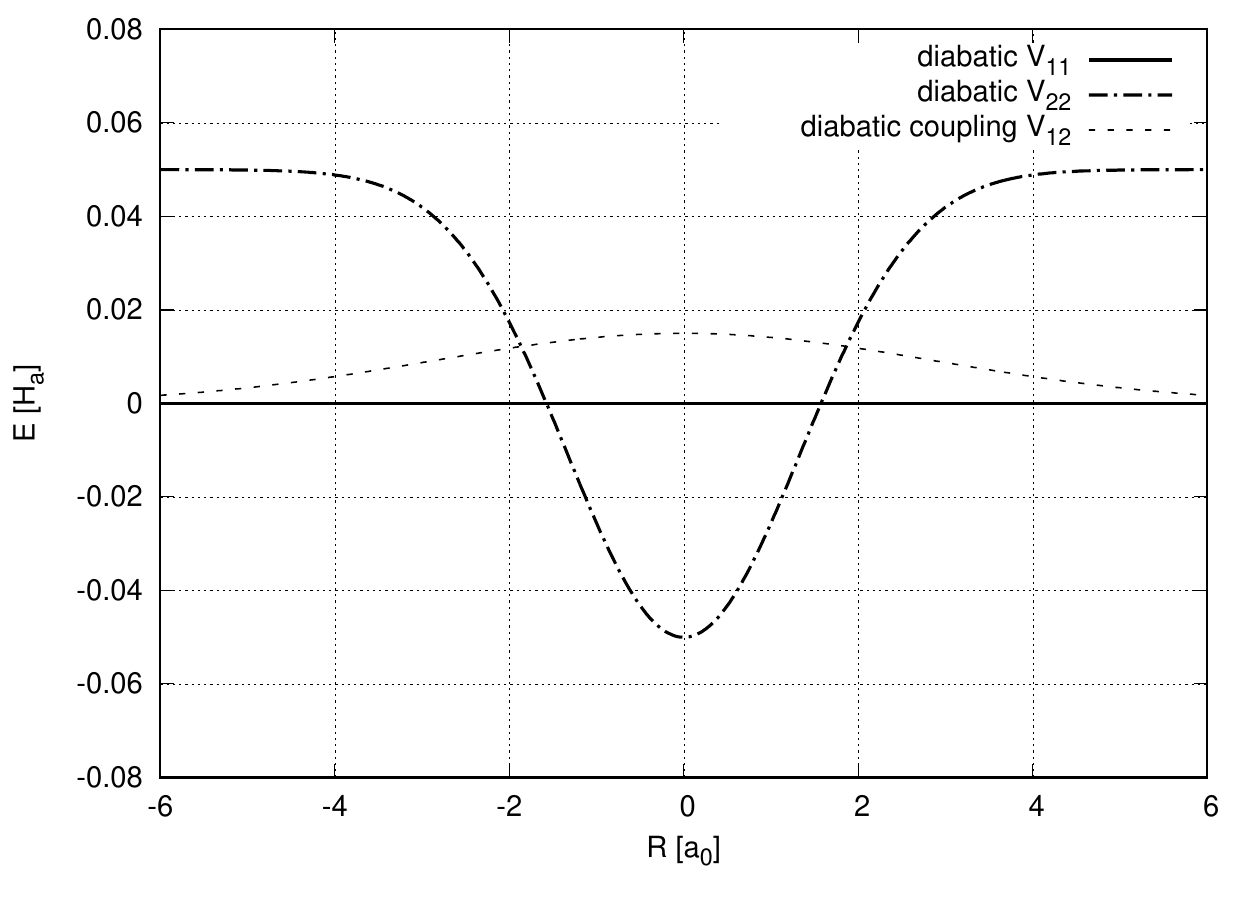}\\\hspace{7mm}(a)}%
    \vtop{\includegraphics[width=\columnwidth]{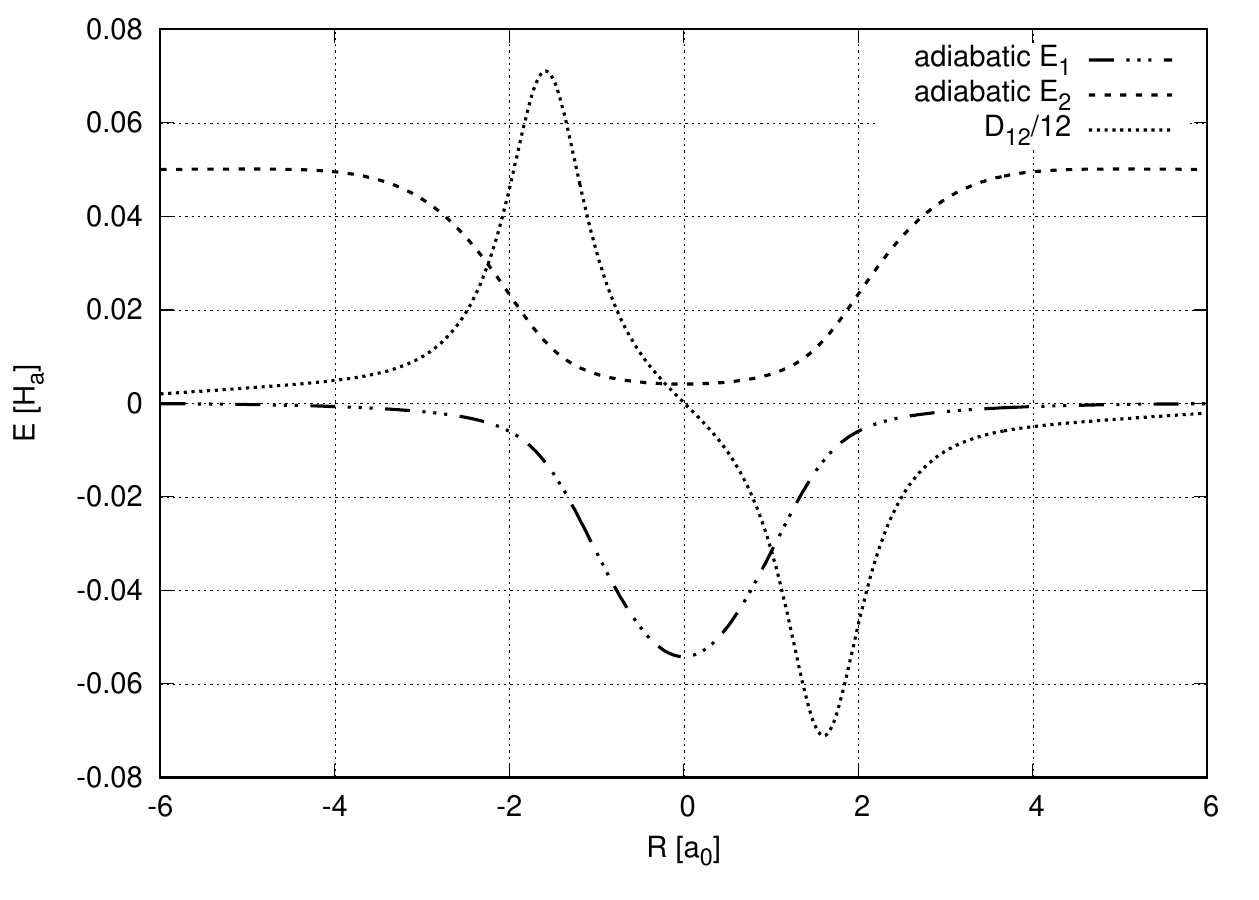}\\\hspace{7mm}(b)}%
    \caption[Model surfaces potential matrix of the dual avoided crossing example]{Model surfaces potential matrix of the dual avoided crossing example;
    (a)~diabatic representation;
    (b)~adiabatic representation, $D_{12}$ is the nonadiabatic coupling element (it is drawn as divided by 12 because the coupling is large).
    }
    \label{fig:potMatrix2}
\end{figure}

\begin{figure}[ht]
    \centering
    \vtop{\includegraphics[width=\columnwidth]{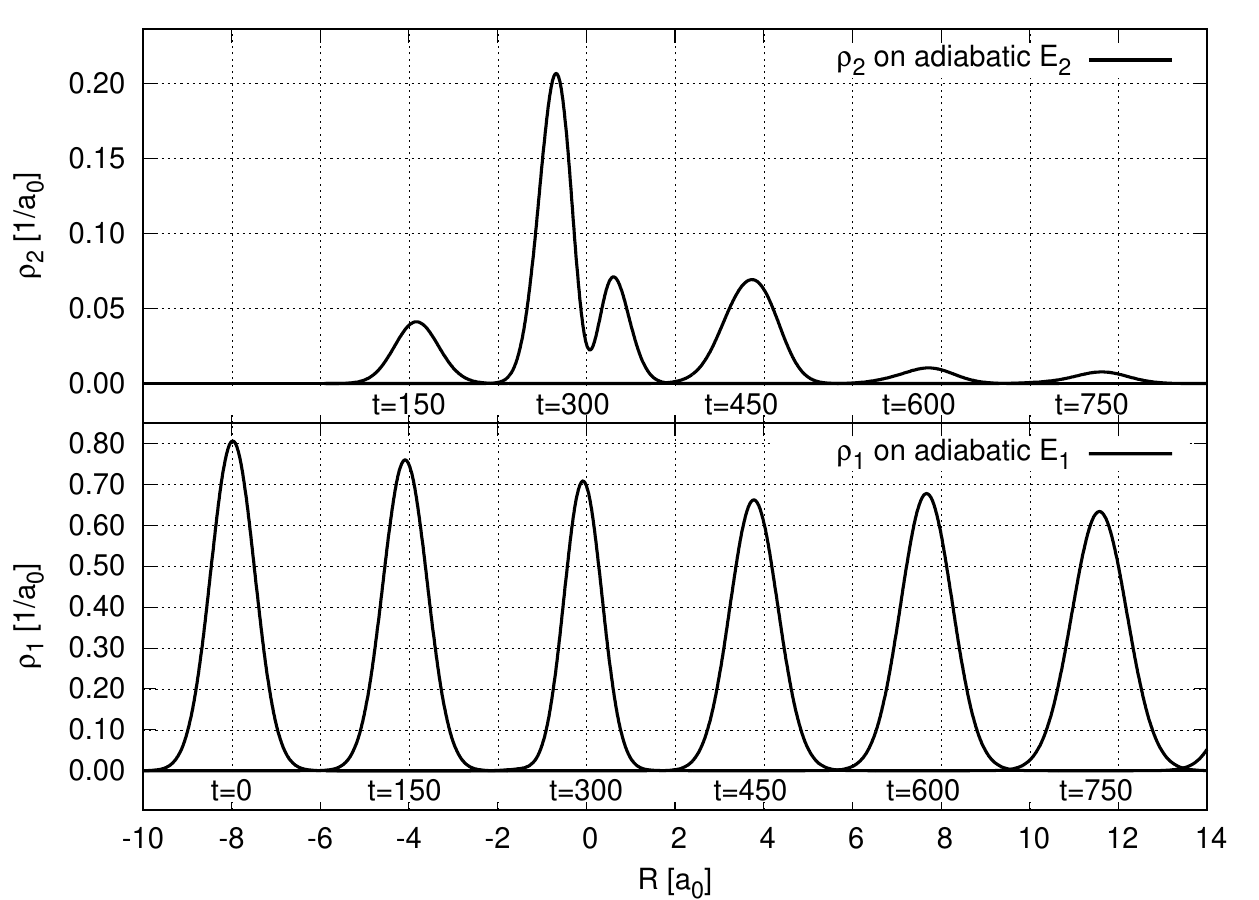}\\\hspace{7mm}(a)}%
    \vtop{\includegraphics[width=\columnwidth]{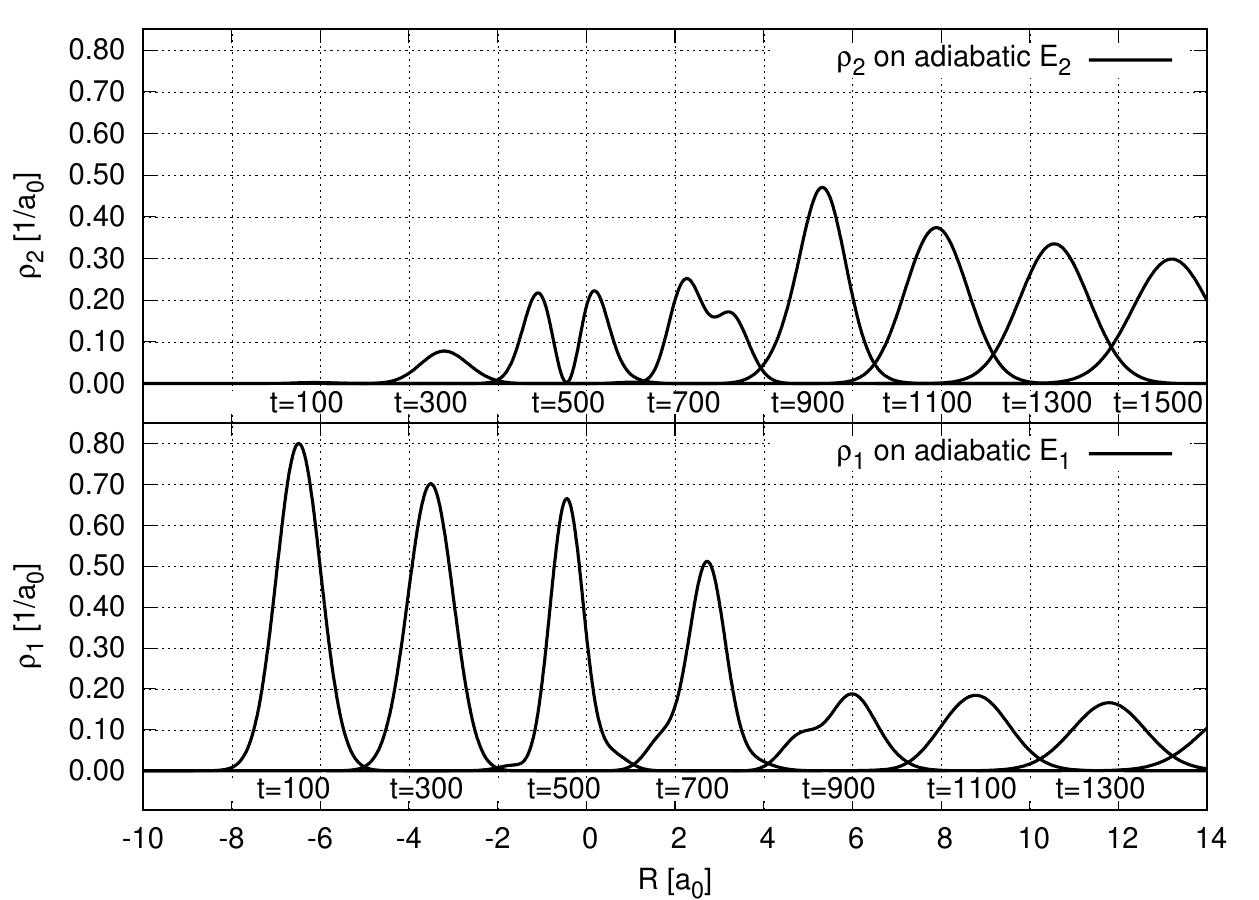}\\\hspace{7mm}(b)}%
    \caption[Evolution of adiabatic probability distributions in dual avoided crossing]{Time evolution of probability distributions on adiabatic energy surfaces, dual avoided crossing;
    (a)~high momentum Gaussian wavepacket $k=52$~a.u.;
    (b)~low momentum Gaussian wavepacket $k=30$~a.u.
    }
    \label{fig:reprNNi14}
\end{figure}

\begin{figure}[ht]
    \centering
    \vtop{\includegraphics[width=\columnwidth]{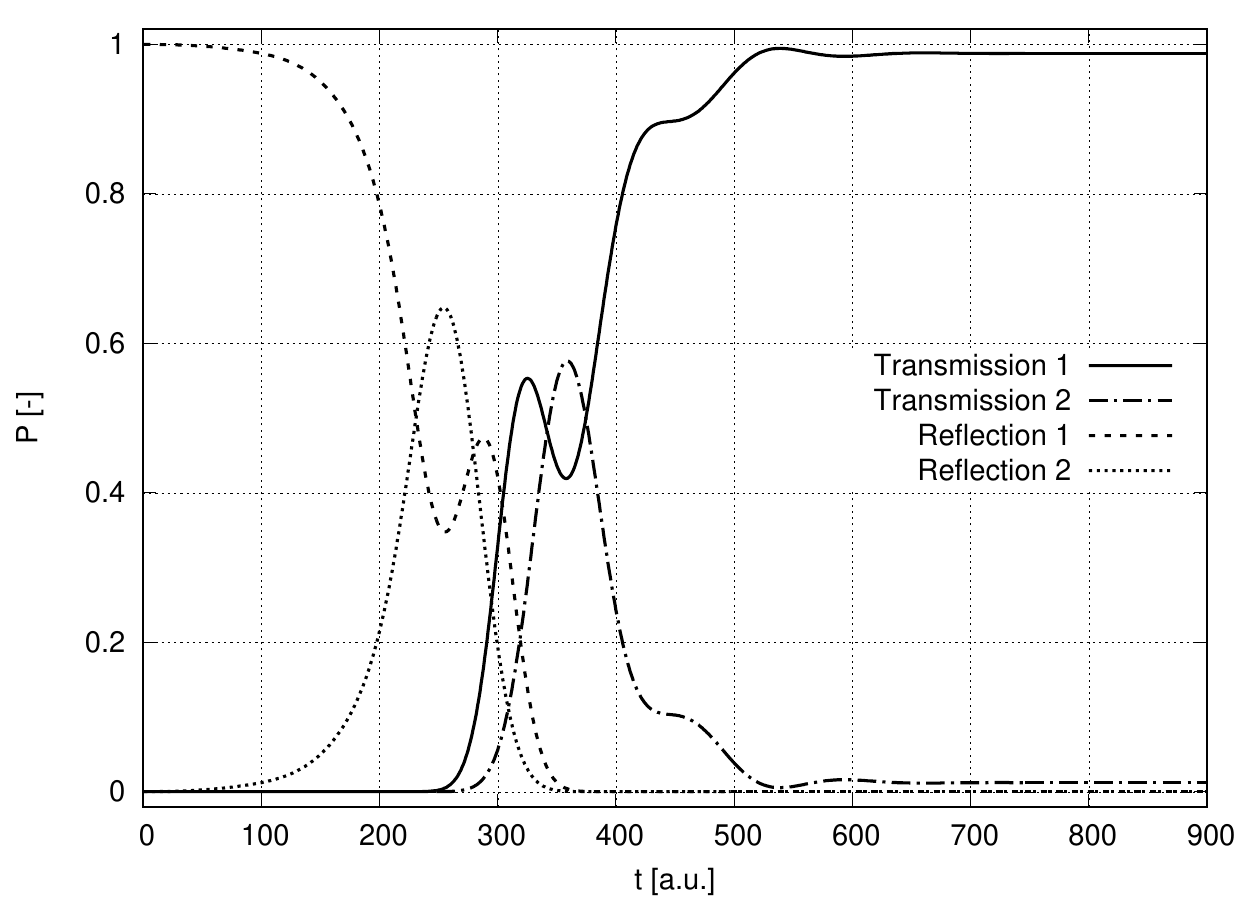}\\\hspace{7mm}(a)}
    \vtop{\includegraphics[width=\columnwidth]{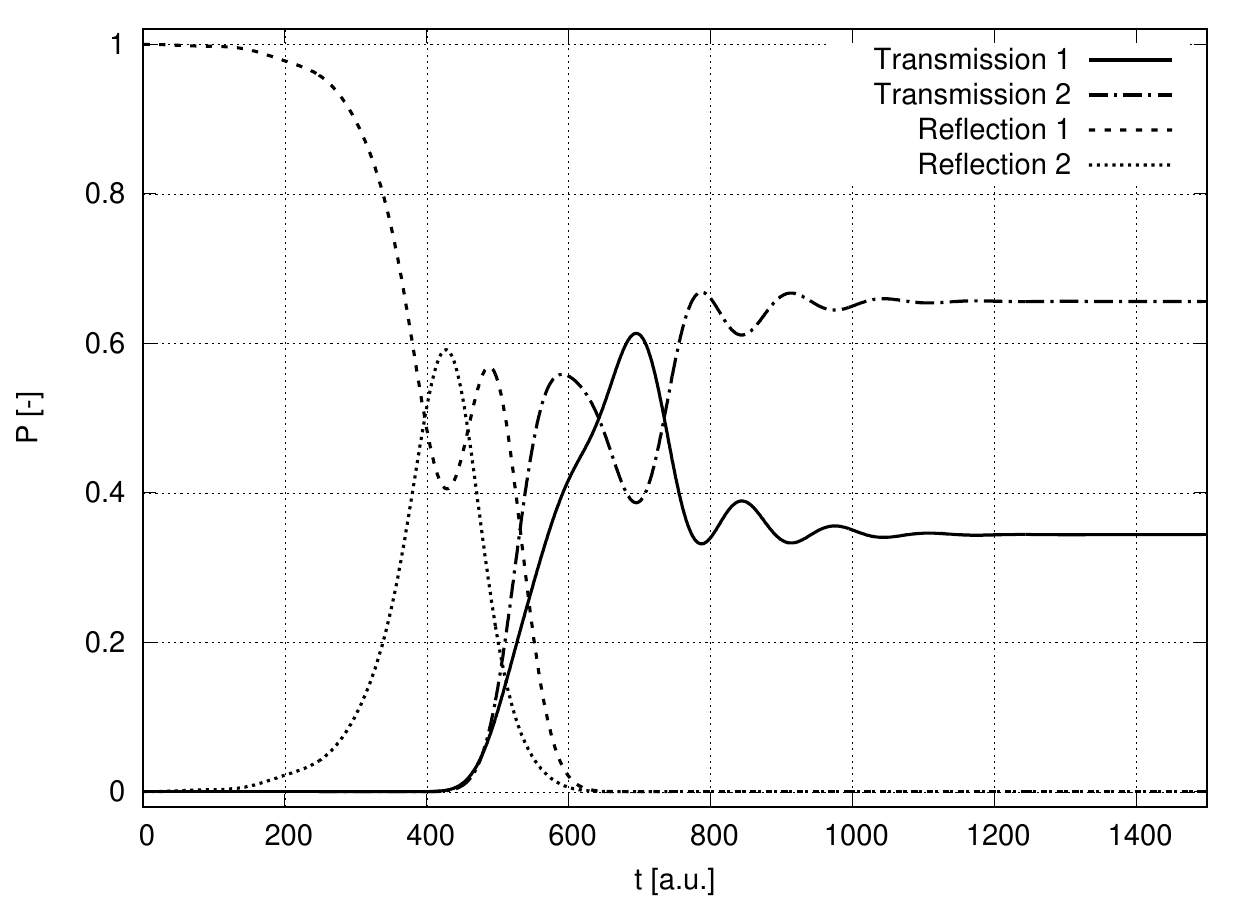}\\\hspace{7mm}(b)}
    \caption[Transmission and reflection probabilities in dual avoided crossing]{Transmission and reflection probabilities as function of time in dual avoided crossing;
    (a)~high momentum Gaussian wavepacket $k=52$~a.u.;
    (b)~low momentum Gaussian wavepacket $k=30$~a.u.
    }
    \label{fig:repr15i16}
\end{figure}

In the second standard benchmark~\cite{Tully1990,Coker1993} the potential
matrix elements in diabatic representation $V_{ij}(R)$ are defined as follows
(\fig{fig:potMatrix2}a):

\begin{equation}
\label{eqDualCrossing}
\begin{split}
V_{11}(R)&=0 \\
V_{22}(R)&=-A e^{-B R^{2}}+E_{0} \\
V_{12}(R)&=V_{21}(R)=C e^{-D R^{2}}
\end{split}
\end{equation}

\noindent
with the parameters assuming values:
$A = 0.1$,
$B = 0.28$,
$C = 0.015$,
$D = 0.06$ and
${E_0=0.05}$.
In this example the diabatic potentials cross each other twice and a wide
Gaussian off-diagonal potential is assumed. The adiabatic surfaces exhibit two
avoided crossings (\fig{fig:potMatrix2}b) and the nonadiabatic coupling element
$D_{12}$ (\eq{eqCoker66}) has two pronounced peaks.

The time evolution of probability distributions is shown on \fig{fig:reprNNi14}
and respective transmission and reflection probabilities are on
\fig{fig:repr15i16}.

The high momentum case in \fig{fig:reprNNi14}a demonstrates the effect of
destructive interference between the first and second crossing.  The wave
packet populates the higher level around $t=150$ ($\rho_2$ in
\fig{fig:reprNNi14}a) then peaks at around $t=350$ (transmission 2~in
\fig{fig:repr15i16}a) and arrives to second crossing at about the same phase at
which it entered the higher level, but this time the nonadiabatic coupling
element $D_{12}(R)$ (\fig{fig:potMatrix2}b) has negative sign.  Same phase of
packet in conjunction with negative sign of $D_{12}(R)$ causes the wave packet
to leave the higher electronic surface almost completely, despite having high
momentum.  This effect can be seen in \fig{fig:repr15i16}a, where in the end
about 98\% of population resides on the lower energy surface.

The low momentum case (\fig{fig:reprNNi14}b) shows constructive interference.
First at around $t=500$ about 32\% of the population enters the higher energy
surface ($\rho_2$ in \fig{fig:reprNNi14}b), then at second crossing additional
32\% enters $\rho_2$. At $t=900$ there is a significant population increase.
\Fig{fig:repr15i16}b shows that about 64\% of the population was transmitted to
the
higher energy surface. This phenomenon is also known as Stükelberg oscillations
and occurs when the time spent by the wave packet between two coupling regions
is an integer or half integer multiple of mean wavepacket oscillations.

Again, the obtained results are in very good agreement with \cite{Tully1990,Coker1993}.

\subsubsection{Precision {and speed} tests}
\label{sec:precision:performance}

\begin{table*}[ht]
\begin{center}
\caption{\revA{Precision test of results from \textit{semi-global} propagator compared against itself, but with higher computation precision. Error is given in terms of ULP (see footnote\textsuperscript{\ref{foot:ULP}}), where the size of ULP for each row is given in the last column (its precise value is in~\Tab{tab:paramsPP}). There is no row for \texttt{boost float128} because there is no result with higher precision against which it can be compared, hence it is used only as reference against which lower precision results are compared.}}
\label{tab:precSemiGlobalSelf}
\begin{tabular}{l l r r l}
\toprule
Benchmark type & Precision      & \texttt{long double} & \texttt{boost float128} & ULP size \\
\midrule
single crossing & \texttt{double}         &   6                  &  6     & $2.2\times10^{-16}$    \\
high $k_0$      & \texttt{long double}    & ---                  &  10    & $1.1\times10^{-19}$ \\
\midrule
single crossing & \texttt{double}         &  14                  &  14    & $2.2\times10^{-16}$    \\
low $k_0$       & \texttt{long double}    & ---                  &  23    & $1.1\times10^{-19}$ \\
\midrule
dual crossing & \texttt{double}         & 14                   &  14    & $2.2\times10^{-16}$    \\
high $k_0$    & \texttt{long double}    & ---                  &  126   & $1.1\times10^{-19}$ \\
\midrule
dual crossing & \texttt{double}         & 7                    &  7     & $2.2\times10^{-16}$    \\
low $k_0$     & \texttt{long double}    & ---                  &  110   & $1.1\times10^{-19}$ \\
\bottomrule
\end{tabular}
\smallskip
\caption{\revA{Precision test of results from \textit{semi-global} propagator (rows in the table) compared against results from global Chebyshev propagator~\cite{Kosloff1984} (columns in the table). Error is given in terms of ULP\textsuperscript{\ref{foot:ULP}}. The size of ULP for each row is given in the last column (its precise value is in~\Tab{tab:paramsPP}).}}
\label{tab:precSemi:Kosloff}
\begin{tabular}{l l r r r l}
\toprule
  & \textit{semi-global} &  \multicolumn{3}{c}{global Chebyshev propagator~\cite{Kosloff1984}} & \\
\cmidrule(rl){2-2}\cmidrule(lr){3-5}
Benchmark type & Precision & \texttt{double} & \texttt{long double} & \texttt{boost float128} & ULP size \\
\midrule
single crossing & \texttt{double}        & 142 &   5          &  6     & $2.2\times10^{-16}$    \\
high $k_0$      & \texttt{long double}   & --- & 3822         &  10    & $1.1\times10^{-19}$ \\
& \texttt{boost float128}& --- & ---          &  548 / 2164$^\ddag$  & $1.9\times10^{-34}$ \\
\midrule
single crossing & \texttt{double}        & 133 &  13          &  14    & $2.2\times10^{-16}$    \\
low $k_0$       & \texttt{long double}   & --- & 1023         &  23    & $1.1\times10^{-19}$ \\
                & \texttt{boost float128}& --- & ---          & 2433 / 7494$^\ddag$  & $1.9\times10^{-34}$ \\
\midrule
dual crossing & \texttt{double}        & 136 &   9          &  14    & $2.2\times10^{-16}$    \\
high $k_0$    & \texttt{long double}   & --- & 15700        & 126    & $1.1\times10^{-19}$ \\
              & \texttt{boost float128}& --- & ---          & 1922   & $1.9\times10^{-34}$ \\
\midrule
dual crossing & \texttt{double}        & 745 &  17          &   7    & $2.2\times10^{-16}$    \\
low $k_0$     & \texttt{long double}   & --- & 20237        & 110    & $1.1\times10^{-19}$ \\
              & \texttt{boost float128}& --- & ---          & 3416   & $1.9\times10^{-34}$ \\
\bottomrule
\end{tabular}
\end{center}$~$\\[-4mm]
\footnotesize{$^\ddag$~Smaller ULP error value is with the number of elements in the series of the global Chebyshev propagator~\cite{Kosloff1984,Kosloff1997} equal to $10\,R$, larger ULP error is with $1.3\,R$, see footnote\textsuperscript{\ref{foot:precR}} for details. Interestingly $1.3\,R$ was enough for \texttt{float128} in dual crossing calculations.}
\end{table*}


\revA{The calculations from two previous sections (single and dual avoided crossings)
are used in this section to compare the precision and performance with the
global Chebyshev propagator~\cite{Kosloff1984,Kosloff1997}.
This section will also demonstrate the speed gain obtained by using the
\textit{semi-global} propagator.}

\revA{The parameters of \textit{semi-global} method used in this comparison are shown
in \Tab{tab:paramsPP}. The $M$ parameter can be very low because there is no
time-dependence in the Hamiltonian, the $K$ parameter was adjusted to be the
minimal value of $K$ where the warning about too small $K$ is not printed (see
\listing{listing4}, line~48). Namely the estimated error of the function of the
matrix is smaller than $\varepsilon$. The $\varepsilon$ parameter is set to the
ULP error\footnote{see footnote\textsuperscript{\ref{foot:ULP}}} because
maximum possible precision in the calculations is used here. And the timestep
used is $\Delta t=1$~a.u.}

\revA{The global Chebyshev propagator~\cite{Kosloff1984,Kosloff1997} has only one parameter,
namely how many elements in the series are calculated, and its value is
actually fixed by the design of the algorithm to be $1.3\,R$ (this
recommendation is given in~\cite{Kosloff1997}), where $R=\frac{\Delta
t}{2\hbar}\left({E}_{max}-{E}_{min}\right)$. This value was found to provide
full precision in nearly all of the calculations, even for higher precision
types\footnote{\label{foot:precR}\revA{To validate this I performed the same
calculation using $10\,R$ and found that the results are the same, except for two
cases which are marked with $^\ddag$ in \Tab{tab:precSemi:Kosloff}}.}.}

\revA{To compare the two algorithms I am performing the same calculation of
transmission and reflection probabilities for all four cases
(\Tab{tab:gaussParams} and Figures \ref{fig:repr11i13} and \ref{fig:repr15i16})
using both \textit{semi-global} method and the global Chebyshev
propagator~\cite{Kosloff1984,Kosloff1997}. The results of transmission and
reflection probabilities are stored in a text file with all significant digits
(column 2~in \Tab{tbl:benchmarked:types}). Since the timestep $\Delta
t=1$~a.u.~and because the transmission and reflection probabilities are stored
for each timestep the text files contain the number of lines equal to the total
simulation duration ("simulation time" in \Tab{tab:gaussParams}) for each of
the four cases. These numbers are then compared against the reference
result and the maximum error in terms of the units of ULP found for
each case is given in
Tables~\ref{tab:precChebySelf}-\ref{tab:precSemi:Kosloff}.}

\revA{The \Tab{tab:precChebySelf} shows the results of comparison of the global Chebyshev
propagator~\cite{Kosloff1984,Kosloff1997} with itself but with higher numerical
precision. The largest error found is for dual crossing, low momentum when
comparing \texttt{long double} with \texttt{float128} precision and equals to
20242 ULP units (where each ULP in this case equals to $1.1\times10^{-19}$).
All ULP errors where a comparison is done between \texttt{long double} and
\texttt{float128} are unusually high and they can be explained by the following
phenomenon: several of the last elements in the series of the Chebyshev
propagator are very small since the Bessel function values are decaying
exponentially, this results in addition of very small values to an otherwise
large values resulting from the previous elements in the series. Hence the
contribution of the last elements in the series vanishes and produces a
slightly less accurate result than would be possible if higher precision was
used.
Very similar ULP error values will appear again in \Tab{tab:precSemi:Kosloff}.}

\revA{The \Tab{tab:precSemiGlobalSelf} shows the results of comparison of
\textit{semi-global} method against itself, but with higher precision. The
largest error is 126 ULP units (where each ULP in this case equals to
$1.1\times10^{-19}$) and is significantly smaller than in
\Tab{tab:precChebySelf} which means that the \textit{semi-global} algorithm
overall has higher accuracy than the global Chebyshev propagator.}

\revA{Finally the \Tab{tab:precSemi:Kosloff} shows the comparison of
\textit{semi-global} algorithm against the global Chebyshev propagator. In this
case it is also possible to compare results for \texttt{float128} precision
between the two algorithms. The largest ULP error is 20237 for the comparison
between \texttt{long double} \textit{semi-global} and \texttt{long double}
global Chebyshev propagator. This value is almost equal to the largest ULP
error found in \Tab{tab:precChebySelf} and indicates that indeed the global
Chebyshev propagator has lower accuracy\footnote{\revA{Because a more accurate
\textit{semi-global} result is compared with less accurate} \revA{global Chebyshev
propagator. Why not the other way around? The answer lies} \revA{in
\Tab{tab:precSemiGlobalSelf} which shows maximum possible errors of the
\textit{semi-global} method} \revA{to be much smaller and also because in
\Tab{tab:precSemi:Kosloff} the largest ULP error between \texttt{long double}
\textit{semi-global} and \texttt{float128} global Chebyshev propagator is 126
ULP units, which indicates that \texttt{long double} in \textit{semi-global}
indeed are accurate.}}. The comparison between \texttt{float128} types of both
methods is more favorable with largest ULP error equal to 3416} \revA{or 7494 if
$1.3\,R$ elements in the series are used in the global Chebyshev propagator.}

\revA{All the calculations discussed in previous paragraphs took a certain amount of
time which was measured to test the computation speed of both algorithms. These
results are summarized in \Tab{tab:perf:Semi:Kosloff} separately for each
precision, averaged over the four simulation cases. The \textit{semi-global}
method turns out to be $1.9\times$ faster
than the global Chebyshev propagator for \texttt{double} precision
and even faster for higher precisions.}



\revA{I would like to point out that the global Chebyshev
propagator~\cite{Kosloff1984,Kosloff1997} is commonly used to validate accuracy
of other time propagation methods~\cite{Leforestier1991}, while in my precision
and speed test it turned out to perform worse than \textit{semi-global}
method in both precision and the calculation speed.}

\revA{The tests discussed in this section are available in the
folder \texttt{Section\_5.2.3\_PrecisionTest} in the supplementary materials.}




\section{Benchmarks of high precision quantum dynamics}
\label{sec:High:Precision}

As mentioned in \Section{sec:Calc:higher}, I would like to emphasize that this
algorithm is implemented in C++ for arbitrary floating point precision types,
specified during compilation.  It works just as well for types with 15, 18 or
33 decimal places.  This is the reason why \texttt{Real} type instead of
\texttt{double} is used e.g.~in line 5~in \listing{listing4}. The \texttt{Real}
type is set during compilation to one of the following types: \texttt{double},
\texttt{long double} or \texttt{float128}. The list of all high precision
types, the number of decimal places and their speed relative to \texttt{double}
is given in \Tab{tbl:benchmarked:types}.  It follows my work on high precision
in YADE, a software for classical dynamics calculations~\cite{Kozicki2022a}.
However the arbitrary precision types \texttt{boost mpfr} and \texttt{boost
cpp\_bin\_float} are currently not available in the quantum dynamics code
because the Fast Fourier Transform (FFT) routines\footnote{used in
\listing{listing3} on page~\pageref{listing3}.} for them are currently
unavailable in the Boost libraries~\cite{Boost2020}.
To resolve this problem I~took part in the Google Summer of Code
2021~\cite{BoostGSoC2021} as a mentor and now the high precision FFT code is in
preparations to be included in the Boost libraries.  The report from high
precision FFT implementation in Boost is available
here~\cite{BoostFFTReport2021}. After that work is complete all types listed in
\Tab{tbl:benchmarked:types} will be available for quantum dynamics
calculations.

\begin{table*}[t]
    \begin{center}
\caption{\revA{Performance comparison of \textit{semi-global} propagator against global Chebyshev propagator~\cite{Kosloff1984}. Values are averaged over all four benchmark types. The \textit{semi-global} propagator is significantly faster than global Chebyshev propagator.}}
\label{tab:perf:Semi:Kosloff}
\begin{tabular}{l r}
\toprule
                        & computation speed of \textit{semi-global} propagator       \\
Precision               & compared to global Chebyshev propagator~\cite{Kosloff1984} \\
\midrule
\texttt{double}         & $1.9\times$  faster \\
\texttt{long double}    & $21.9\times$ faster \\
\texttt{boost float128} & $8.1\times$  faster \\
\bottomrule
\end{tabular}
\smallskip
    \caption[The high-precision benchmark: comparison between classical dynamics~\cite{Kozicki2022a} and quantum dynamics.]{The high-precision benchmark: comparison between classical dynamics~\cite{Kozicki2022a} and quantum dynamics. The speed is shown as relative to speed at \texttt{double} precision}
    \label{tbl:benchmarked:types}
        \begin{tabular}{l r r r}
            \toprule
            ~                              & Decimal & Classical dynamics          & Quantum dynamics      \\
            Precision                      & places  & speed w.r.t \texttt{double}~\cite{Kozicki2022a} & speed w.r.t \texttt{double} \\
            \toprule
            \texttt{float}                 & 6       & $1.01\times$ faster         &     \\
            \texttt{double}                & 15      & ---                         & ---,  $\varepsilon = 2.2\times10^{-16}$ \\
            \texttt{long double}           & 18      & $1.4\times$ slower          & $3.5\times$ slower, $\varepsilon = 1.1\times10^{-19}$ \\
            \texttt{boost float128}        & 33      & $4.7\times$ slower          &  $50\times$ slower, $\varepsilon = 2.2\times10^{-16}$ \\
            \texttt{boost float128}        & 33      &                             & $170\times$ slower, $\varepsilon = 1.9\times10^{-34}$ \\
            \texttt{boost mpfr}$^\dag$     & 62      & $13.5\times$ slower         &     \\
            \texttt{boost mpfr}            & 150     & $19.1\times$ slower         &     \\
            \texttt{boost cpp\_bin\_float} & 62      & $24.2\times$ slower         &     \\
            \bottomrule
        \end{tabular}
    \end{center}$~$\\[-4mm]
\footnotesize{$^\dag$~for future comparison with \texttt{libqd-dev} library.}\\
\end{table*}


\Tab{tbl:benchmarked:types} shows the speed comparison between different
high-precision types, relative to \texttt{double}, separately for classical
dynamics and quantum dynamics\footnote{
\revA{It should be noted that the two types of problems used in the
benchmark:~quantum vs.~classical dynamics have entirely different characteristics
and should not be compared \textit{per se}. The only meaningful conclusion by comparing
the two columns in \Tab{tbl:benchmarked:types} is that the excessive times are different.
This conclusion cannot be generalized}.}.
The classical dynamics are reproduced from YADE benchmark~\cite{Kozicki2022a}
to serve as a reference.
I have done the quantum dynamics benchmark using the example of atom in an
intense laser field from previous \Section{sec:validate:atom}. I used the same
parameters with the exception of changing $\varepsilon$ (\eq{eqSemiVII}) to
match the ULP error\footnote{see
footnote\textsuperscript{\ref{foot:ULP}} on page \pageref{foot:ULP} for
details.} of selected precision as provided in table.

The \texttt{long double} precision in quantum dynamics is about $3.5\times$
slower and in exchange provides about $2000$ times greater precision.  This
might be useful in some situations for quick verification of results.

For the \texttt{float128} type I did the benchmark twice, first for the
$\varepsilon$ same as for \texttt{double} type, then for the significantly
smaller $\varepsilon$ matching the \texttt{float128} type.
We can see that using $\varepsilon = 2\times10^{-16}$ from
\texttt{double} for the
\texttt{float128} makes it about $50\times$ slower\footnote{%
\revA{The \textit{semi-global} algorithm using \texttt{float128}
with $\varepsilon = 2\times10^{-16}$ is} performing the same amount of mathematical
operations as if the \texttt{double} type was used.}.
When using full \texttt{float128} precision then the decreased error tolerance
$\varepsilon$ forces more iterations in \eq{eqSemi62a} consequently making it
$170\times$ slower.
If one wished to calculate with larger tolerance, say $\varepsilon =
1\times10^{-25}$, then still \texttt{float128} has to be used, and it will have
speed somewhere between the two values in \tab{tbl:benchmarked:types}.
%



I would like to mention that from my experience~\cite{Kozicki2022b}, it is better to increase
$\Delta t$ and allow more sub-iterations in \Eq{eqSemi62a} as it speeds up
calculation more than changing the $K$ and $M$ parameters.  For example, this
atom in the laser field calculation took 30~seconds with \texttt{double} and
$\Delta t=0.1$~a.u.~and 10~minutes with \texttt{double} and $\Delta
t=0.0041(6)$~a.u. both having the same $\varepsilon = 2\times10^{-16}$ error
tolerance.

Moreover, the original code~\cite{Schaefer2017}, which deals with a single
Schrödinger equation, was written in Matlab~\cite{Quarteroni2014} and the same
simulation which took me 30~seconds in C++, took about 5~minutes in
Octave\footnote{Octave is an open-source version of
Matlab~\cite{Quarteroni2014}.
\revA{The speed gain might be smaller in comparison to Matlab, since it is known
to be faster than Octave}.
\revA{However} I have no access to commercial Matlab software \revA{to quantify the difference}.
Single core was used in both cases, C++ and Octave.}.  So the $10 \times$ speed
gain due to migrating from Octave to C++ can now be wisely spent on higher
precision calculations or on simulating larger systems.

\section{Accompanying C++ source code package}
\label{sec:code:package}


The C++ source code has been run and tested on Linux Ubuntu, Debian and
Devuan\footnote{The tests were run on: Devuan Chimaera, Debian Bullseye, Debian
Bookworm, Ubuntu 20.04 and Ubuntu 22.04. Older linux distributions couldn't work due
to too old version of \revStyle{the} libeigen library.}, and it
should run without significant tweaks on any modern GNU/Linux operating system.
Some tweaks might be needed for other
operating systems such as MacOS or Windows. In general the source code is very
minimal in the sense that there is only the implemented \textit{semi-global}
algorithm in files \texttt{SemiGlobal.hpp} and \texttt{SemiGlobal.cpp} together
with the tests listed in \Section{sec:validation} which can be invoked by
various \texttt{makefile} calls (see \texttt{makefile} for full list),
such as \texttt{make plot\_Coker} or \texttt{make plot\_Atom}.

All available make targets can be
readily found by reading the comments inside the \texttt{makefile}. One notable
target is \texttt{make testAllFast} which does almost all the tests and takes
about 20 minutes. The full test \texttt{make testAll} takes about 7 hours, due
to \texttt{float128} precision being significantly slower.
From all implemented tests, the file \texttt{test\_single\_dual\_crossing.cpp}
is the most interesting because it contains code for working with multiple
electronic levels as well as the placeholder code for time dependence in the
potential (although it is not used%
\footnote{The time dependent potential is used in
\texttt{test\_atom\_laser\_ABC.cpp} and \texttt{test\_source\_term.cpp}. A calculation of a full
multi-level system with time dependent potential is in preparations to be
published in a separate paper, where we investigate the dynamics of a three
level NaRb system subject to a laser excitation~\cite{Kozicki2023a}.}).
It is this source code with a generic Hamiltonian \texttt{calc\_Hpsi}
(\Listing{listing5}) which is
discussed in \Section{chapter:Time:Dependent}.  Other test files
\texttt{test\_source\_term.cpp} and \texttt{test\_atom\_laser\_ABC.cpp} are the
direct translations of Matlab/Octave code from \cite{Schaefer2017}.

The library dependencies are following:
\texttt{libfftw3-dev}     \cite{Frigo2005} (version $\geq$ 3.3.8),
\texttt{libboost-all-dev} \cite{Boost2020} (version $\geq$ 1.71.0) and
\texttt{libeigen3-dev}    \cite{Eigen2020} (version $\geq$ 3.3.7-2).

\revB{See file \texttt{README.pdf} in the accompanying source code package for additional
details about how to use the \textit{semi-global} algorithm with custom Hamiltonian in custom
coordinate representation, such as curvilinear coordinates. The section \textit{Usage}
in the \texttt{README.pdf} will be maintained and improved as questions arise from users.}

The \texttt{SemiGlobal.hpp} and \texttt{SemiGlobal.cpp} files are written to be
self contained in a generic way adhering to C++ coding standards. This means
that these files are readily available to use in other C++ projects with only
minimal changes at the interface to ``glue'' the code to a different codebase.
Depending on whether high precision calculations are required in the other
software package the file \texttt{Real.hpp} might be used as well.

\section{Conclusions}
\label{sec:Conclusions}

In this paper, I present an implementation of the \textit{semi-global}
algorithm for coupled Schrödinger equations with the time-dep\-endent Hamiltonian
and nonlinear inhomogeneous source term as a means to describe the time-dependent
processes in femto- and attosecond chemistry.
%
The code works for multiple coupled electronic states and supports high precision
computations with types \texttt{long double} (18 decimal places) and
\texttt{float128} (33 decimal places). Higher arbitrary precision types
will become available in the future once FFT algorithm for them becomes
ready to use in the C++ boost library.

The \textit{semi-global} algorithm is verified to work correctly by comparing
its results with five reference solutions: (1) atom in an intense laser field
(2) single avoided crossing (3) dual avoided crossing (4) Gaussian packet in a forced
harmonic oscillator and (5) forced harmonic oscillator with an inhomogeneous
source term. All of them are available in the accompanying source code package.

\revA{The precision and performance test revealed that the \textit{semi-global}
algorithm is more accurate than the global Chebyshev propagator, while being
about two times faster for \texttt{double} precision and even faster for higher
precisions.}

This C++ code, which is $10\times$ faster than the original Matlab/Octave code upon
which it was based, can be used to produce reference high accuracy solutions of
various problems such as: nonlinear problems, problems with inhomogeneous
source term, mean field approximation, Gross-Pitaevskii approximation or
scattering problems.

The attached C++ source code is self contained which makes it possible to reuse
the algorithm in different software packages.


\section*{Supplementary materials}

This paper is accompanied by a file \texttt{SemiGlobalCpp.zip}
which contains the C++ source code for GNU/Linux described above, licensed
under the GNU GPL v.2 license.

\section*{Author Information}
\noindent{\textbf{Corresponding Author}}\\
\noindent * E-mail: \href{mailto:jkozicki@pg.edu.pl}{jkozicki@pg.edu.pl}\\[1mm]
\noindent{\textbf{ORCID}}\\
\newcommand{\OrcidLink}[1]{\href{https://orcid.org/#1}{#1}}
\noindent Janek Kozicki:        \OrcidLink{0000-0002-8427-7263}\\[1mm]
\noindent{\textbf{Conflicts of interest}}\\
There are no conflicts of interest to declare.

\section*{Acknowledgments}

This publication is based upon work from COST Action AttoChem, CA18222
supported by COST (European Cooperation in Science and Technology).
I would like to thank Józef E. Sien\-kie\-wicz, Patryk Jasik and Tymon Kilich
for fruitful discussions on this work. Additionally, I would like to thank
Ido Schaefer, Hillel Tal-Ezer and Ronnie Kosloff for their original idea
of \textit{semi-global} algorithm and its implementation in Matlab.
Also, I would like to thank the two anonymous reviewers for their insight
and suggestions.

\bibliography{manuscript.bib}
\end{document}